\documentclass[12pt]{article}
\usepackage{jheppub}

\usepackage{amsmath,bbm,array,amsfonts,graphicx,wrapfig,lscape,float,multirow,longtable,rotating,subfigure,epstopdf}
\usepackage{float}

\newcommand{\be}{\begin{equation}}
\newcommand{\ee}{\end{equation}}
\newcommand{\beq}{\begin{equation}}
\newcommand{\beql}[1]{\begin{equation}\label{#1}}
\newcommand{\eeq}{\end{equation}}
\newcommand{\ba}{\begin{array}}
\newcommand{\ea}{\end{array}}
\newcommand{\bea}{\begin{eqnarray}}
\newcommand{\beal}[1]{\begin{eqnarray}\label{#1}}
\newcommand{\eea}{\end{eqnarray}}
\newcommand{\ben}{\begin{enumerate}}
\newcommand{\een}{\end{enumerate}}
\newcommand{\bean}{\begin{eqnarray*}}
\newcommand{\eean}{\end{eqnarray*}}
\newcommand{\eref}[1]{(\ref{#1})}
\newcommand{\sref}[1]{\S\ref{#1}}

\newcommand{\fref}[1]{Figure \ref{#1}}
\newcommand{\btab}[1]{\begin{tabular}{#1}}
\newcommand{\etab}{\end{tabular}}

\newcommand{\comment}[1]{}

\newcommand{\IC}{\mathbb{C}}

\newcommand{\qed}{\nobreak \ifvmode \relax \else
      \ifdim\lastskip<1.5em \hskip-\lastskip
      \hskip1.5em plus0em minus0.5em \fi \nobreak
      \vrule height0.75em width0.5em depth0.25em\fi}

\def\beqa{\begin{eqnarray}}
\def\eeqa{\end{eqnarray}}

\newcolumntype{C}[1]{>{\centering\arraybackslash}m{#1}}

\newcommand{\IS}{{\bf S}}
\newcommand{\NN}{{\cal{ N}}}
\newcommand{\IZ}{{\mathbb{Z}}}

\def\makeatletter{\catcode`\@=11}
\makeatletter
\def\mathbox#1{\hbox{$\m@th#1$}}%
\def\math@ccstyles#1#2#3#4#5#6#7{{\leavevmode
     \setbox0\mathbox{#6#7}%
     \setbox2\mathbox{#4#5}%
     \dimen@ #3%
     \baselineskip\z@\lineskiplimit#1\lineskip\z@
     \vbox{\ialign{##\crcr
            \hfil \kern #2\box2 \hfil\crcr
            \noalign{\kern\dimen@}%
            \hfil\box0\hfil\crcr}}}}
\def\mathaccstyles{\math@ccstyles\maxdimen}
\def\maththroughstyles{\math@ccstyles{-\maxdimen}}
\def\unity%
{\maththroughstyles{.45\ht0}\z@\displaystyle {\mathchar"006C}\displaystyle 1}

\title{Charting Class $\mathcal{S}_k$ Territory}

\author[a,b]{Sebasti\'an Franco,}
\author[c]{Hirotaka Hayashi}
\author[c]{and Angel Uranga}

\affiliation[a]{
Physics Department, The City College of the CUNY \\
160 Convent Avenue, New York, NY 10031, USA}

\affiliation[b]{The Graduate School and University Center, The City University of New York  \\
365 Fifth Avenue, New York NY 10016, USA }

\affiliation[c]{Instituto de F\'isica Te\'orica IFT-UAM/CSIC \\
C/ Nicol\'as Cabrera 13-15, Universidad Aut\'onoma de Madrid, 28049 Madrid, Spain
}

\emailAdd{sfranco@ccny.cuny.edu,h.hayashi@csic.es,angel.uranga@uam.es}

\abstract{
We extend the investigation of the recently introduced class ${\cal S}_k$ of 4d $\NN=1$ SCFTs, by considering a large family of quiver gauge theories within it, which we denote $\mathcal{S}^1_k$. These theories admit a realization in terms of $\IZ_k$ orbifolds of Type IIA configurations of D4-branes stretched among relatively rotated sets of NS-branes. This fact permits a systematic investigation of the full family, which exhibits new features such as non-trivial anomalous dimensions differing from free field values and novel ways of gluing theories. We relate these ingredients to properties of compactification of the 6d (1,0) superconformal ${\cal T}_N^k$ theories on spheres with different kinds of punctures. We describe the structure of dualities in this class of theories upon exchange of punctures, including transformations that correspond to Seiberg dualities, and exploit the computation of the superconformal index to check the invariance of the theories under them.
}

\preprint{
\begin{flushright}IFT-UAM/CSIC-15-036\end{flushright} 
}

\begin{document}

\maketitle


\section{Introduction}

A very fruitful approach to the study of superconformal field theories (SCFTs) in various dimensions has been their definition in terms of some underlying geometric or combinatorial object. 

In recent years, a successful incarnation of this general program has been to consider compactifications of 6d SCFTs (on Riemann surfaces, 3-manifolds, etc) to engineer SCFTs in various dimensions and with different amounts of SUSY. The most prominent example of this construction is the class $\mathcal{S}$ of 4d $\mathcal{N}=2$ SCFTs, which corresponds to compactifications of the 6d (2,0) theory on punctured Riemann surfaces \cite{Gaiotto:2009we, Gaiotto:2009hg} (see \cite{Gaiotto:2014bja} for a review). This construction leads to both Lagrangian and non-Lagrangian theories, which are connected by S-dualities that translate into transformations of the underlying Riemann surface. Certain weak coupling regimes can be realized in terms of Hanany-Witten like brane configurations \cite{Hanany:1996ie} of D4-branes suspended among NS5-branes \cite{Witten:1997sc}.

The class $\mathcal{S}$ has been extended by introducing the $\mathcal{N}=1$ gluing, generalizing the way in which elementary building blocks are combined into more complicated theories \cite{Maruyoshi:2009uk, Benini:2009mz,Tachikawa:2011ea, Bah:2011je,  Bah:2011vv,  Bah:2012dg, Beem:2012yn, Gadde:2013fma, Maruyoshi:2013hja, Bah:2013aha, Agarwal:2013uga, Agarwal:2014rua, Giacomelli:2014rna, McGrane:2014pma} (see \cite{Xie:2013gma, Bonelli:2013pva, Xie:2013rsa, Yonekura:2013mya, Xie:2014yya} for their Hitchin systems). The theories admit weak coupling limits described in terms of D4-branes suspended between mutually rotated NS5-branes \cite{Elitzur:1997fh}. They are also related to compactifications of the 6d $(2,0)$ theory on Riemann surfaces with two kinds of minimal punctures.

An even broader $\mathcal{N}=1$ generalization, denoted class $\mathcal{S}_k$, was recently proposed in \cite{Gaiotto:2015usa}. It is defined as the compactification of the 6d (1,0) SCFTs $\mathcal{T}_k^N$\footnote{On the other hand, the compactification of the 6d (1,0) theory on a torus yields the class $\mathcal{S}$ theory \cite{Ohmori:2015pua}. }, which arise from $N$ M5-branes at an $A_{k-1}$ orbifold singularity in M-theory, over punctured Riemann surfaces. In \cite{Gaiotto:2015usa}, an interesting family of quiver gauge theories within this class was engineered by starting from a Type IIA configuration of parallel NS5-branes and $N$ transverse D4-branes sitting at an $A_{k-1}$ singularity, like those introduced in \cite{Lykken:1997gy}. Before the orbifold action, this setup preserves $\mathcal{N}=2$ in 4d, therefore the resulting $\NN=1$ theories can be constructed as $\IZ_k$ orbifolds of $\NN=2$ class $\mathcal{S}$ theories. 

With the goal of deepening our understanding of general $\mathcal{S}_k$ theories, in this paper we embark in the exploration of a wider family of quivers in this class, which includes those that were the primary focus of \cite{Gaiotto:2015usa}. We denote these theories class $\mathcal{S}^1_k$. The superindex emphasizes the fact that the theories are $\NN=1$ even before the $\IZ_k$ orbifold involved in the 6d  $\mathcal{T}_k^N$ theory. An interesting feature of class $\mathcal{S}^1_k$ is that it admits a Type IIA embedding in terms of non-parallel NS5-branes (usually denoted NS- and NS'-branes) and D4-branes preserving 4d $\NN=1$ SUSY, over an $A_{k-1}$ singularity compatible with the same SUSY preserved by the branes. This fact enables a general and systematic analysis of the full family. Its investigation reveals, in very general terms, new features of $\mathcal{S}_k$ theories such as the interplay between Seiberg duality and the 6d picture, the importance of non-trivial anomalous dimensions and novel ways of gluing theories.

The theories we study play an analogue role to the one of linear quivers for class $\mathcal{S}$. Remarkably, like the quivers studied in \cite{Gaiotto:2015usa}, class $\mathcal{S}^1_k$ theories are a simple subset with cylinder topology of the general family of Bipartite Field Theories (BFTs) \cite{Franco:2012mm,Xie:2012mr}. BFTs are 4d $\NN=1$ quiver gauge theories (including theories with enhanced $\NN=2,4$) that are defined by bipartite graphs on Riemann surfaces, possibly including boundaries (for additional works on BFTs, see e.g. \cite{Franco:2012wv,Heckman:2012jh,Franco:2013ana,Hanany:2012vc}).\footnote{In another interesting development, the same bipartite graphs have recently appeared in the on-shell diagram formalism for scattering amplitudes in 4d $\NN=4$ SYM \cite{ArkaniHamed:2012nw,Arkani-Hamed:2014bca,Franco:2015rma}.} BFTs contain and generalize brane tilings, which correspond to bipartite graphs on a 2-torus. They have been extensively studied and play a prominent role in the study of the 4d $\NN=1$ SCFTs that arise on the worldvolume of stacks of D3-branes at the tip of toric CY threefold singularities \cite{Hanany:2005ve,Franco:2005rj,Franco:2005sm,Butti:2005sw}.

This paper is organized as follows. In \sref{sec:general-class} we construct class $\mathcal{S}^1_k$. We also analyze how the exchange of punctures is translated into Seiberg duality. In \sref{sec:global-symm} we introduce a graphical prescription for determining anomaly-free global symmetries for these theories. Their marginal deformations are studied in \sref{sec:marginal}, where we find agreement with the expectations coming from realizing them as compactifications on Riemann surfaces from 6d. In \sref{section_theories_from_building_blocks} we discuss how to engineer new theories by gluing a pair of them along maximal punctures or closing minimal punctures. The superconformal index for our theories is studied in \sref{sec:index}. We conclude in \sref{section_conclusions}.

\bigskip

\section{$\mathbb{Z}_k$ Orbifold of $\mathcal{N}=1$ Linear Quivers}
\label{sec:general-class}

In this paper we introduce a new class of 4d $\mathcal{N}=1$ SCFTs, which we call class $\mathcal{S}^1_k$,  and which fall in the class of $\mathcal{S}_k$ described by compactifications of 6d (1,0) SCFTs $\mathcal{T}_k^N$ on punctured Riemann surfaces $\Sigma$. Our theories differ from the main class considered in \cite{Gaiotto:2015usa} in that they admit two basic kinds of minimal punctures. In other words, the embedding of the Riemann surface is twisted such that the parent theory (on the covering space of the orbifold) preserves $\NN=1$ in our class $\mathcal{S}_k^1$ rather than $\NN=2$ in \cite{Gaiotto:2015usa}. The class $\mathcal{S}_k^1$ thus provides a more complete perspective of the whole set of $\mathcal{S}_k$ theories obtainable from the  6d (1,0) $\mathcal{T}_k^N$ theory.

The class $\mathcal{S}^1_k$ is defined in terms of quiver gauge theories, which we can build in two stages. First, in this section we construct them as $\IZ_k$ orbifolds of certain $\mathcal{N}=1$ linear quivers. They are analogous to $\mathcal{N}=2$ linear quivers for class $\mathcal{S}$ \cite{Gaiotto:2009we} and the class of $\mathcal{N}=1$ cylindrical quivers recently introduced in \cite{Gaiotto:2015usa}. These theories admit a Type IIA brane realization and have a cylindrical topology, and correspond to compactifications of 6d (1,0) SCFTs $\mathcal{T}_k^N$ on punctured spheres. Following the terminology of \cite{Gaiotto:2015usa}, we will also refer to these theories as {\it core theories} in class $\mathcal{S}^1_k$. Starting from them, other class ${\cal S}_k$ theories can be achieved via higgsing, as mentioned in \sref{section_closing_minimal_punctures}.

\bigskip

\subsection{The Core Theories and Their Type IIA Brane Realizations}

We will now construct  class $\mathcal{S}^1_k$ quiver theories, which are defined in terms of Type IIA brane configurations of D4-branes stretched among mutually rotated NS5-branes \cite{Elitzur:1997fh} (building on \cite{Hanany:1996ie}). Specifically, we consider D4-branes along the directions 0123 and stretching along 4 between NS5-branes along the directions 012356 and rotated NS5-branes (denoted NS'-branes) along 012378. The construction of SCFT is achieved by considering an equal number $N$ of D4-branes suspended in each interval bounded by NS- or NS'-branes, as well as in the semi-infinite regions at the ends. 

\begin{figure}[h]
\begin{center}
\includegraphics[width=14cm]{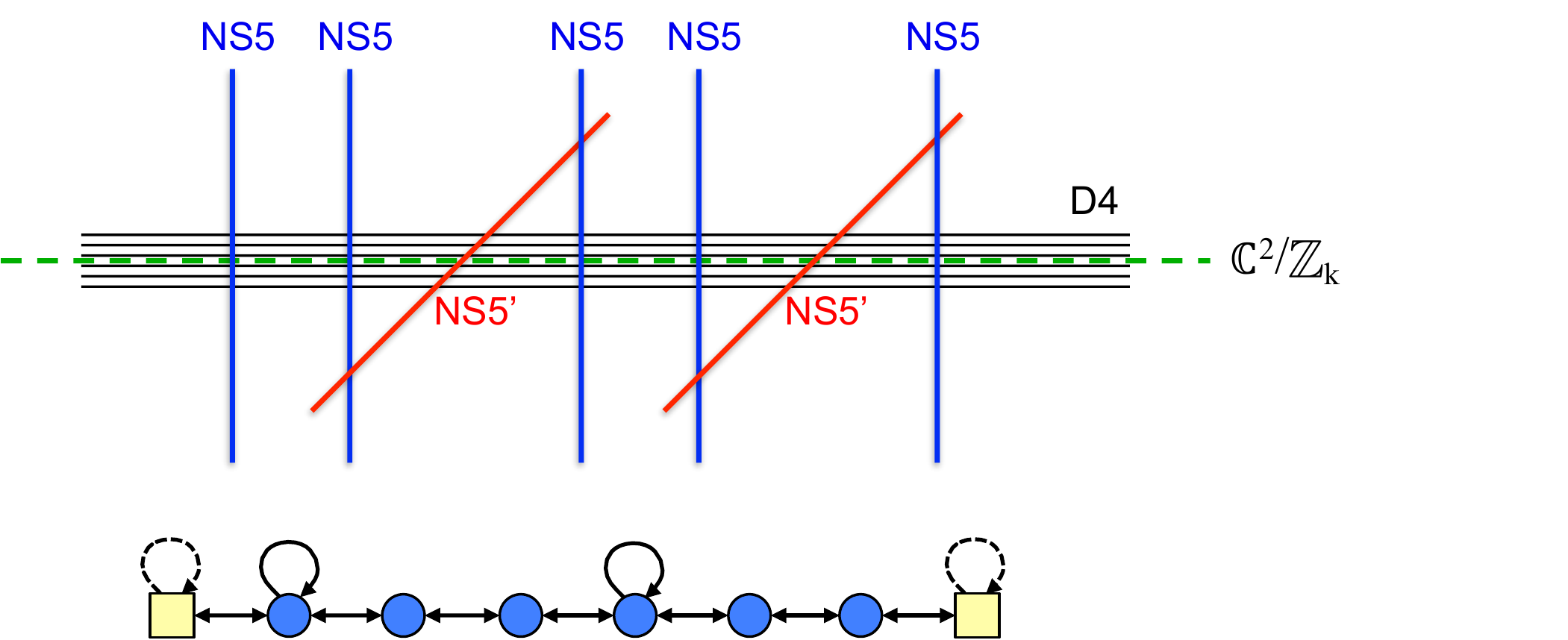}
\caption{Type IIA brane configuration for core class $\mathcal{S}^1_k$ theories. The configuration involves D4-branes suspended among $n_5$ NS5 and $n_5'$ NS5'-branes, in an arbitrary order, and located at a $\IC^2/\IZ_k$ orbifold. At the bottom we show the associated linear quiver corresponding to the brane configuration {\it before} the $\mathbb{Z}_k$ orbifold. Blue circular nodes correspond to gauge symmetries while yellow square nodes correspond to global symmetries. Dashed arrows represent non-dynamical fields in the adjoint representation of the global nodes.}
\label{basic_IIA}
\end{center}
\end{figure}

The orbifold theories are obtained upon quotienting the directions $z=x^5+ix^6$ and $w=x^7+ix^8$ by the $\IZ_k$ action $(z,w)\to (e^{2\pi i/k}z,e^{-2\pi i/k}w)$. A typical example of this kind of Type IIA configuration is shown in \fref{basic_IIA}. Since much information of the orbifold theory is encoded in the structure of the parent linear quiver theory (shown at the bottom of \fref{basic_IIA}, we explain the basic features of the latter. 

\begin{figure}[h]
\begin{center}
\includegraphics[width=7.5cm]{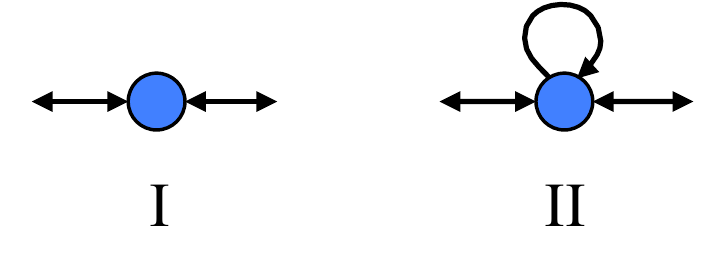}
\caption{The two basic types of nodes in the linear quiver. Type I corresponds to D4-branes stretched between an NS5-NS5' pair. Type II corresponds to D4-branes stretched between an NS5-NS5 or an NS5'-NS5' pair. }
\label{Type_I_and_II_nodes}
\end{center}
\end{figure}

A linear quiver theory realized in terms of $n_5$ NS-branes and $n_5'$ NS'-branes has a total number of gauge groups $n:=n_5+n_5'-1$, and two endpoint global symmetry groups. We label them collectively with an index $i=0,1,\ldots,n,n+1$. We indicate gauge and global $SU(N)$ symmetry groups with circular and square nodes in the quiver, respectively. Gauge groups are classified into two types, Type I and Type II, according to what kind of D4-brane interval they arise from, equivalently according to their corresponding node in the parent theory, see Figure \ref{Type_I_and_II_nodes}. 

\medskip

$\bullet$ A Type I node corresponds to $N$ D4-branes stretched between an NS-NS' pair, and contains an $SU(N)$ gauge factor,\footnote{\label{note-u1}The $U(1)$'s that would naively arise from the $U(N)$ in the D-brane realization are actually massive by the brane bending mechanism in \cite{Witten:1997sc}. From the field theory perspective, they can also be argued to be absent in the SCFT because they are IR free. In any picture, they remain as anomaly-free global symmetries of the configuration.} bifundamental chiral multiplets $X_{i,i\pm 1}$, $X_{i\pm 1,i}$, and a quartic superpotential interaction among them, given (modulo sign) by
\footnote{In superpotential terms, and in forthcoming definitions of mesons, e.g. (\ref{mesons}), traces of the monomials are implicit throughout the paper.} 
\beqa
W_{I}=X_{i,i-1}X_{i-1,i}X_{i,i+1}X_{i+1,i}.
\eeqa 

\smallskip

$\bullet$ A Type II node corresponds to $N$ D4-branes stretched between an NS-NS or an NS'-NS' pair, and contains an $SU(N)$ gauge factor,  bifundamental chiral multiplets $X_{i,i\pm 1}$, $X_{i\pm 1,i}$, one adjoint chiral multiplet $\Phi_i$, and a cubic superpotential (modulo sign)
\beqa
W_{II}= X_{i-1,i}\Phi_iX_{i,i-1}-X_{i+1,i}\Phi_i X_{i,i+1}.
\eeqa

\smallskip

Type II nodes preserve an $\NN=2$ supersymmetry, which is ultimately broken in other sectors; actually, if $n_5=0$ or $n_5'=0$, all nodes are Type II and we recover the standard $\NN=2$ Type IIA brane configurations \cite{Witten:1997sc}. We denote $\tilde{n}$ the number of the Type I nodes.

The global symmetry nodes, i.e. nodes 0 and $n+1$, correspond to the semi-infinite stacks of D4-branes at both sides of the brane setup. For concreteness, let us consider node 0. The semi-infinite D4-branes can move in the 56 or 78 planes, depending on whether the first NS5-brane is an NS- or an NS'-brane. In both cases, this motion results in a mass term for the bifundamental fields stretching between nodes 0 and 1. We can thus translate this motion into a chiral field $\Phi_0$ transforming in the adjoint representation of node 0 with a superpotential coupling $W_0=\Phi_0 X_{0,1} X_{1,0}$. Since $\Phi_0$ has a 5d support, it is a non-dynamical field from the 4d viewpoint. From now on, we indicate such non-dynamical fields as dashed arrows in the quiver diagram. The same analysis applies to node $n+1$, for which motion of the corresponding branes can be translated into a non-dynamical adjoint $\Phi_{n+1}$.

\begin{figure}[h]
\begin{center}
\includegraphics[width=7.5cm]{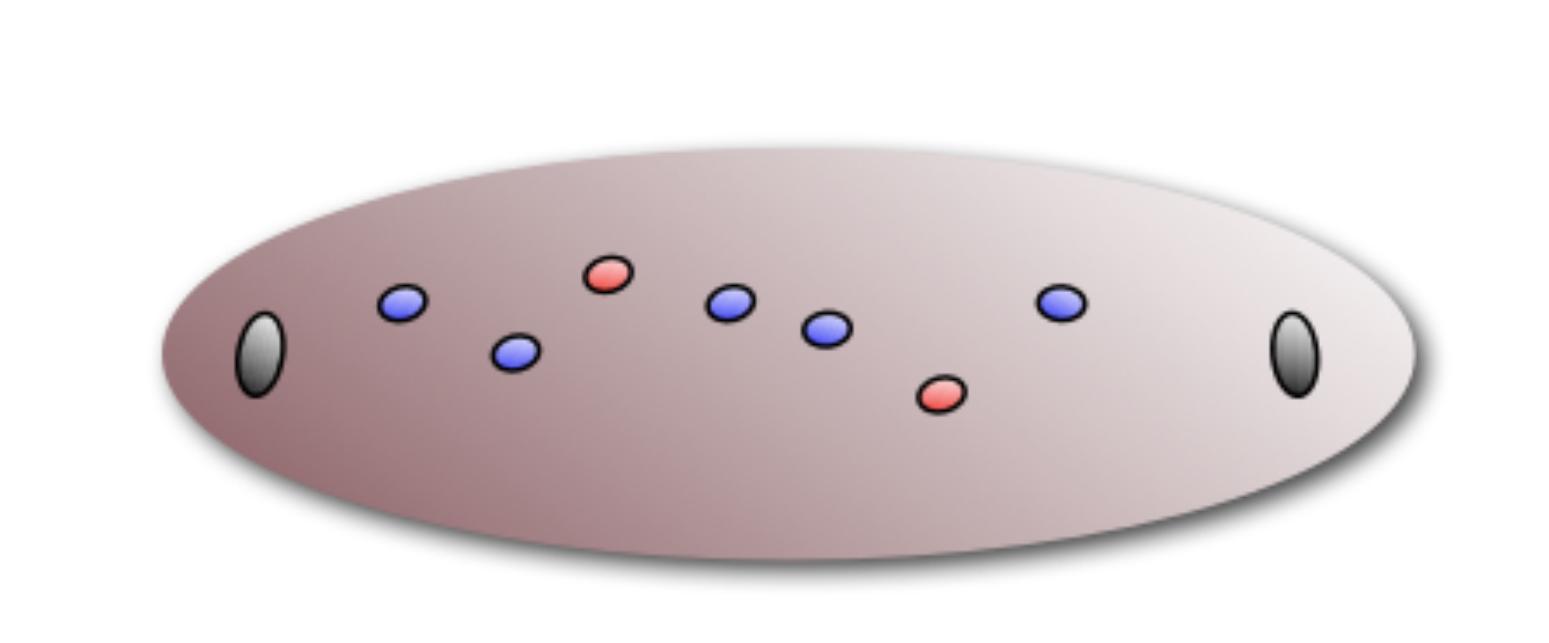}
\caption{Riemann surface for the theory in Figure \ref{basic_IIA}. Grey punctures are maximal, and the blue and red punctures correspond to the two kinds of minimal ones.}
\label{riemann}
\end{center}
\end{figure}

The orbifold by $\IZ_k$ can be carried out in field theory using ideas in \cite{Uranga:1998vf}, or at the level of the Type IIA brane configuration by extending the results in \cite{Lykken:1997gy}; it is actually straightforward using the bipartite graphs (aka dimer diagrams) to be introduced soon.\footnote{To be precise, the theories with all nodes being $SU(N)$ as described below are obtained by taking $SU(kN)$ nodes in the parent theory, with the $k$-fold multiplicity associated to the regular representation of the $\IZ_k$ orbifold group.} The resulting $\NN=1$ theories are weak coupling limits of compactification of the 6d $(1,0)$ $\mathcal{T}_k^N$ SCFT on a Riemann surface given by a sphere with two maximal punctures and $(n+1)$ minimal punctures of two different kinds (corresponding to the location of NS- and NS'-branes), see \fref{riemann}. The dictionary between these punctures and the global symmetries of the theory will be developed in \sref{sec:global-symm}.
%

The result of the orbifold is that after the $\mathbb{Z}_k$ orbifold, each gauge node produces a column with $k$ $SU(N)$ nodes,\footnote{\label{note-u1k}The $U(1)$ in the $U(N)$ that naively appear in the D-brane realization are generically anomalous and massive, with anomaly cancelled by a Green-Schwarz mechanism, discussed in \cite{Ibanez:1998qp} in a T-dual realization. Non-anomalous combinations correspond to $U(1)$'s inherited from the parent theory; as explained in footnote \ref{note-u1}, they disappear as gauge symmetries, but remain as anomaly-free global symmetries.} and the quiver results in a tiling of a cylinder topology. We can then label the gauge nodes as $(i, a)$ with $i=1, \ldots, n, a=1, \ldots, k$. Again we have two types of gauge nodes depending on whether they descend from Type I and Type II nodes in the parent theory. We recover the models in \cite{Gaiotto:2015usa} for $n_5=0$ or $n_5'=0$, i.e. when all nodes descend from Type II nodes. Nodes descending from a Type I node have $2N$ fundamental chiral multiplets and $2N$ anti-fundamental chiral multiplets (with the multiplicity $N$ actually corresponding to a fundamental or anti fundamental representation under some near-neighbouring node). Nodes descending from a Type II node have $3N$ fundamental chiral multiplets and $3N$ anti-fundamental chiral multiplets. We have $\tilde{n}k$ Type I nodes and $(n-\tilde{n})k$ Type II nodes. 

Up to an overall sign, each of these nodes comes with the following contributions to the superpotential
\begin{eqnarray}
W_I & = & X_{(i-1, a)}^{(i, b)}X_{(i, a-1)}^{(i-1, b)} X_{(i+1, b)}^{(i, a-1)} X_{(i, a)}^{(i+1, b)}, \label{superI}\\
W_{II} & = & \phi_{(i, a)}^{(i, a-1)} X_{(i-1, b)}^{(i, a)}X_{(i, a-1)}^{(i-1, b)} - \phi_{(i, a)}^{(i, a-1)} X_{(i+1, b)}^{(i, a)}X_{(i, a-1)}^{(i+1, b)}, \label{superII}
\end{eqnarray}
where $b=a$ or $a-1$. $X^{(i, a)}_{(j, b)}$ and $\phi^{(i, a)}_{(j, b)}$ indicates a chiral multiplet corresponding to an arrow from $SU(N)_{(i, a)}$ to $SU(N)_{(j, b)}$. We have $\tilde{n}k$ quartic couplings and $2(n-\tilde{n})k$ cubic couplings. 

\begin{figure}[h]
\begin{center}
\includegraphics[height=7cm]{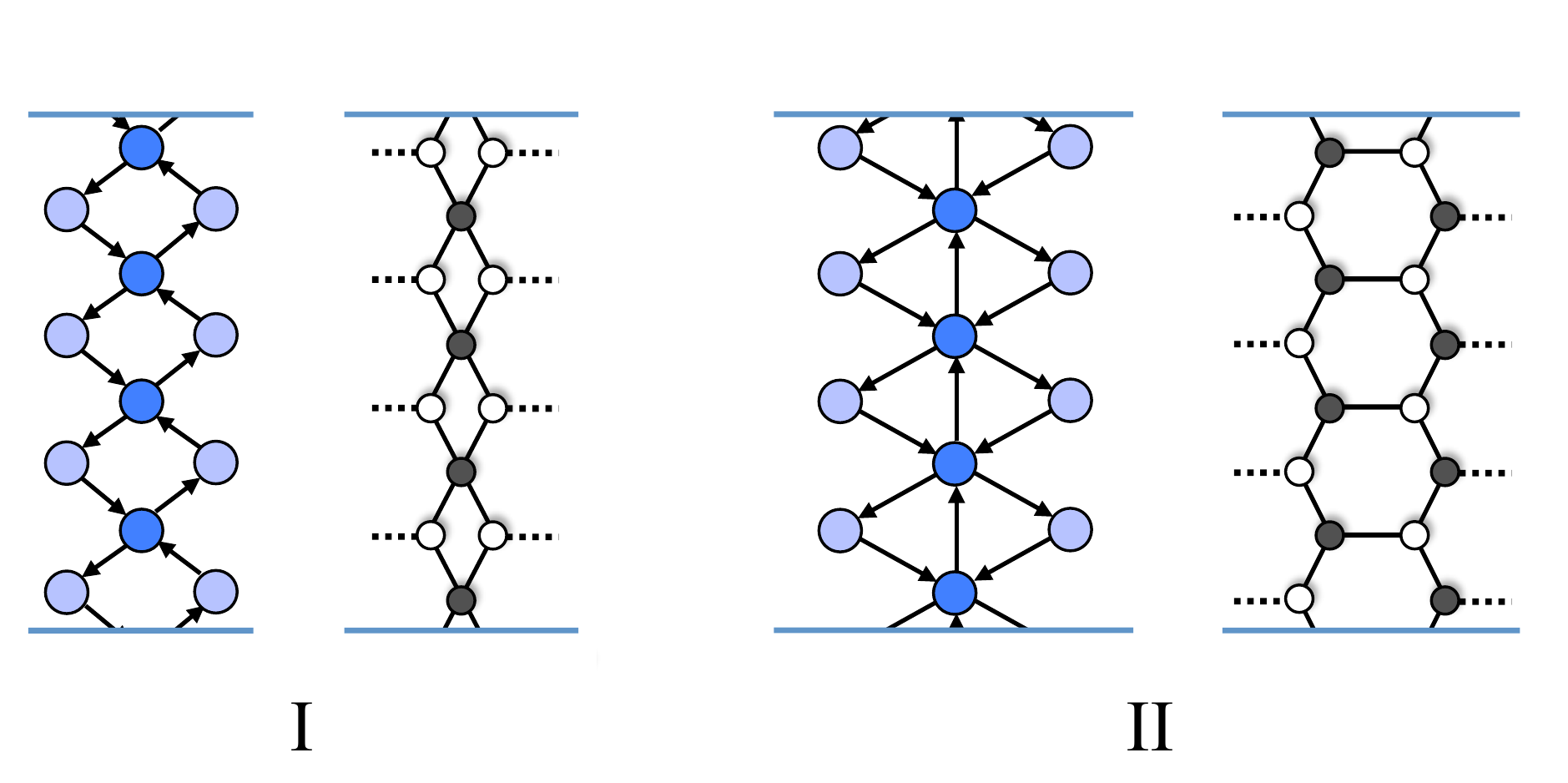}
\vspace{-.2cm}\caption{The two possible kinds of columns of nodes descending from parent Type I or Type II nodes after orbifolding, in quiver and dimer languages. The top and bottom blue lines are identified, giving the quiver and dimer a cylinder topology.} 
\label{orbifold-nodes}
\end{center}
\end{figure}

The basic structure of the two kinds of columns of nodes is shown in \fref{orbifold-nodes}. In these quivers, every oriented plaquette corresponds to a term in the superpotential.

Remarkably, all the core theories in our construction belong to the general class of {\it bipartite field theories} (BFTs) introduced in \cite{Franco:2012mm,Xie:2012mr}. BFTs are 4d $\mathcal{N}=1$, quiver gauge theories (also including theories with enhanced $\NN=2,4$) whose Lagrangian is specified by a bipartite graph on a Riemann surface, which might contain punctures. For core theories, the Riemann surface is a cylinder. We refer the reader to \cite{Franco:2012mm} for a thorough discussion of BFTs. In order to be self-contained, we summarize the dictionary connecting bipartite graphs (or dimers, for short) to BFTs in Table \ref{tab:dictionaryBFT}. In \fref{orbifold-nodes}, we also present the dimers associated to the two types of columns.

\begin{table}[h]
\begin{center}
\begin{tabular}{|p{.25\textwidth}|p{.70\textwidth}|}
\hline
{\bf Graph} & {\bf BFT} \\
\hline
\hline
Internal Face  & $SU(N)$ gauge symmetry group with $n N$ flavors. \\
($2n$ sides) & \\
\hline
External Face & $SU(N)$ global symmetry group \\
\hline
Edge between faces $i$ and $j$ & Chiral superfield in the bifundamental representation of 
groups $i$ and $j$ (or the adjoint representation if $i=j$). The chirality, i.e.\ orientation, of the bifundamental is such that it goes clockwise around white nodes and counter-clockwise around black nodes. External legs correspond to non-dynamical fields.
\\
\hline
$k$-valent internal node & Superpotential term made of $k$ chiral superfields. Its sign is $+/-$ for a white/black node, respectively. \\
\hline
\end{tabular}
\end{center}
\caption{The dictionary relating bipartite graphs on Riemann surfaces to BFTs.}
\label{tab:dictionaryBFT}
\end{table}

\begin{figure}[h]
\begin{center}
\includegraphics[height=7cm]{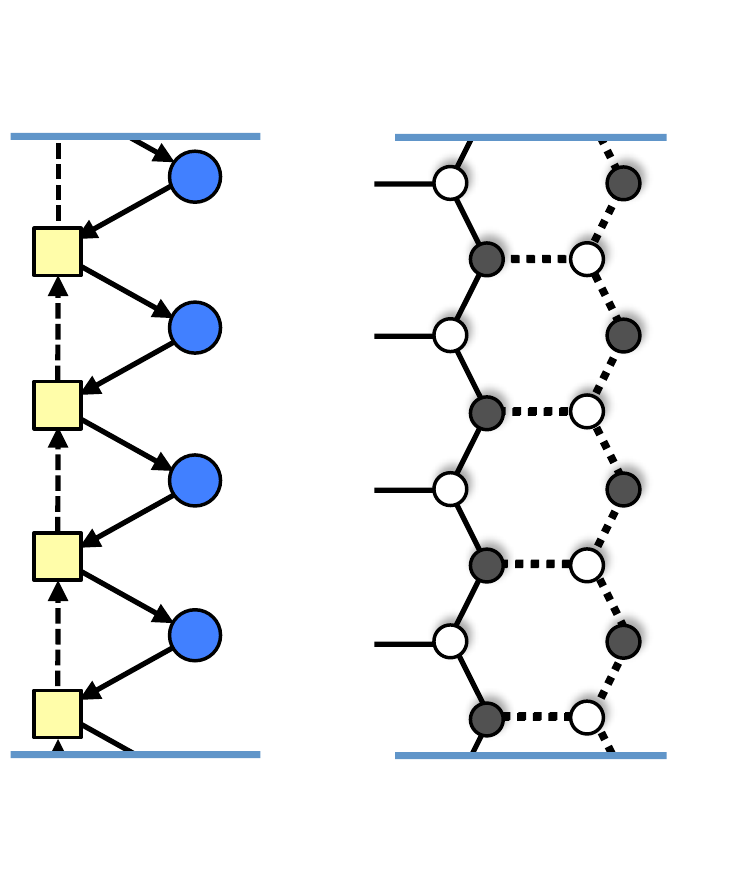}
\vspace{-.9cm}\caption{The result of orbifolding the global symmetry nodes in the parent theory, in quiver and dimer languages. The non-dynamical dashed arrows endow the maximal puncture with an orientation, in this case ascending.} 
\label{external_nodes_orb}
\end{center}
\end{figure}

Let us now consider the $\IZ_k$ orbifold action on the global symmetry nodes. For concreteness, let us focus on node 0; an identical discussion applies to node $n+1$. The orbifold turns node 0 into $k$ $SU(N)$ global nodes. In addition, the non-dynamical adjoint $\Phi_0$ gives rise to $k$ non-dynamical bifundamental fields $\phi_{(0, a)}^{(0, a-1)}$ between pairs of global nodes. They are coupled to chiral bifundamental fields connected to the first column of gauge nodes through cubic superpotential terms of the form
\beq
W_{0} = \phi_{(0, a)}^{(0, a-1)} X_{(1, b)}^{(0, a)}X_{(0, a-1)}^{(1, b)} .
\eeq
For non-zero vevs of the fields $\phi_{(0, a)}^{(0, a-1)}$, these terms are relevant deformations of the SCFT and trigger RG flows. Unless we indicate otherwise, we will thus freeze the vevs of all $\phi_{(0, a)}^{(0, a-1)}$'s to zero. Accordingly, we will not consider these terms when analyzing the marginal deformations in \sref{sec:marginal}. The corresponding column in the quiver and bipartite graph are shown in \fref{external_nodes_orb}.

Even though the $\phi_{(0, a)}^{(0, a-1)}$ are non-dynamical, there are various reasons for why it is useful to keep them in our discussion: they keep track of possible marginal deformations of the SCFTs, they endow maximal punctures with a natural orientation as shown in \fref{external_nodes_orb}, and they turn into dynamical fields when gluing punctures, as we will discuss in \sref{sec:gluing-maximal}. 

In the coming section, we will combine the basic building blocks we have discussed above into full theories.

\bigskip

\subsection{The Full Theories: Two Standard Orderings}
\label{sec:two-orderings}

In this section we put together the ingredients we previously developed to construct the core quiver gauge theories that are candidates for describing the compactification of the 6d (1,0) $\mathcal{T}_k^N$ theory on a cylinder with minimal punctures of two types. The two types of punctures arise from the two possible orientations of the NS5-branes in Type IIA and of the corresponding M5-branes in the M-theory lift. There are $n_5$ and $n_5'$ punctures of each type.

In the Riemann surface picture, the maximal punctures correspond to the semi-infinite D4-branes, which carry certain global symmetries of the SCFT, and minimal punctures correspond to the NS- and NS'-branes. One potentially surprising feature in the correspondence is that there is no notion of ordering of the minimal punctures in $\Sigma$, whereas there is an ordering of the NS- and NS'-branes in the Type IIA configurations. In this section we present two standard or canonical orderings, but eventually argue that the orderings are actually physically unimportant, since the SCFTs obtained from different orderings are related by Seiberg dualities \cite{Seiberg:1994pq}, as we explain in more detail in \sref{sec:seiberg}. 

\begin{figure}[h]
\begin{center}
\includegraphics[width=14cm]{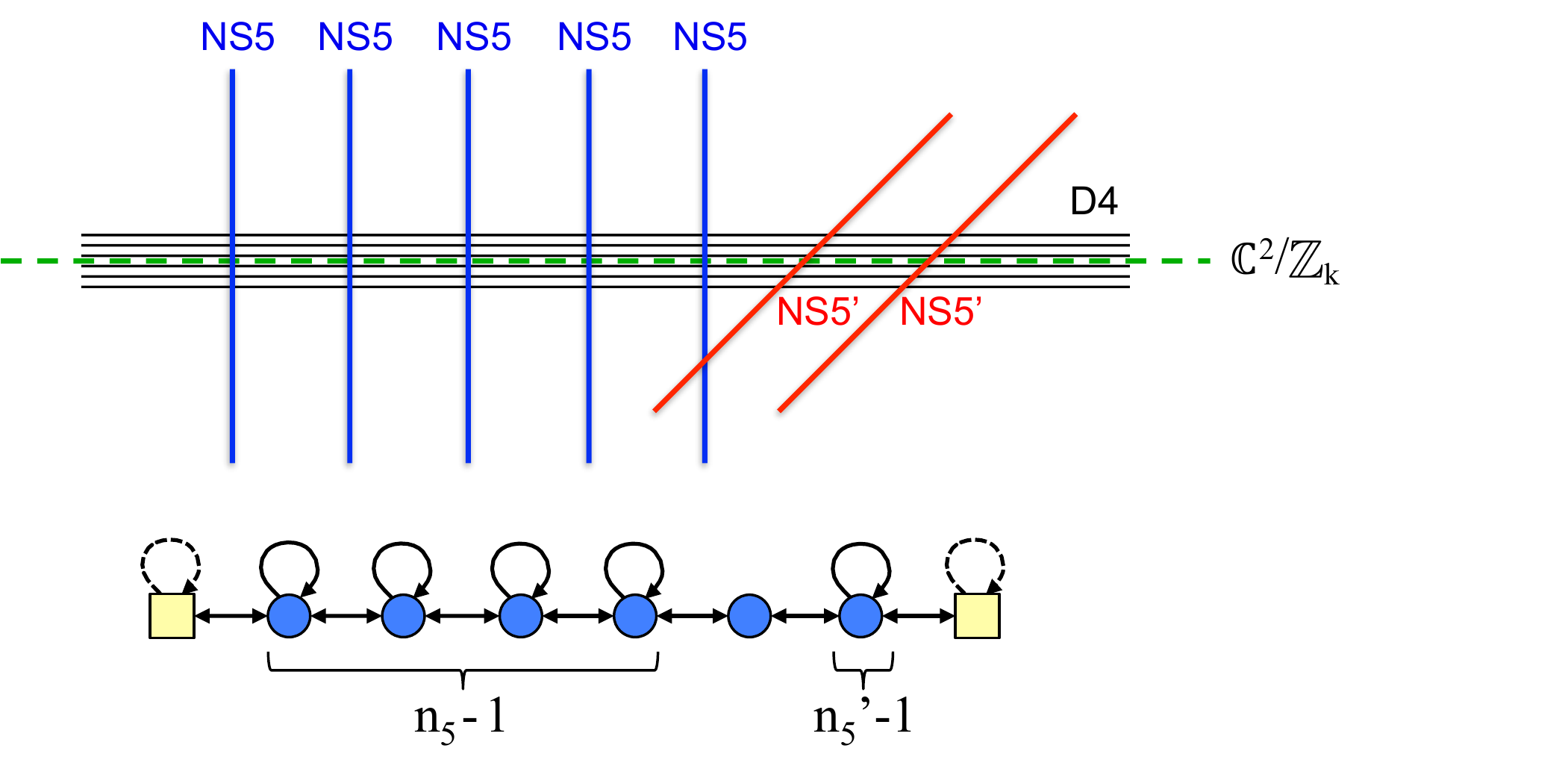}
\caption{Top: Type IIA brane configuration corresponding to a core theory with $n_5=5$ and $n_5'=2$ in the first standard ordering. Bottom: the associated linear quiver {\it before} the $\mathbb{Z}_k$ orbifold.}
\label{IIA_quiver_ordering_1}
\end{center}
\end{figure}

Our theories are realized as Type IIA brane configurations with $n_5$ NS- and $n_5'$ NS'-branes respectively. Without loss of generality we can take $n_5\geq n_5'$. These branes are ordered according to their positions in the direction 4 (along which the D4-brane stacks are suspended). A standard ordering of the branes is to locate first the different $n_5$ NS-branes, and then the $n_5'$ NS'-branes. This left-right distribution of NS- and NS'-branes is shown in \fref{IIA_quiver_ordering_1}. Before the orbifold, the configuration describes two sectors, preserving (different) $\NN=2$ supersymmetries (vector-like bifundamentals coupled to adjoints by cubic superpotentials), joined by a gauge sector preserving the common $\NN=1$ supersymmetry (no adjoint and quartic superpotential among the vector-like bifundamentals). In other words, the two sectors consist of $n_5-1$ and $n_5'-1$ Type II nodes, respectively, and are separated by a single Type I node.

\begin{figure}[h]
\begin{center}
\includegraphics[width=14cm]{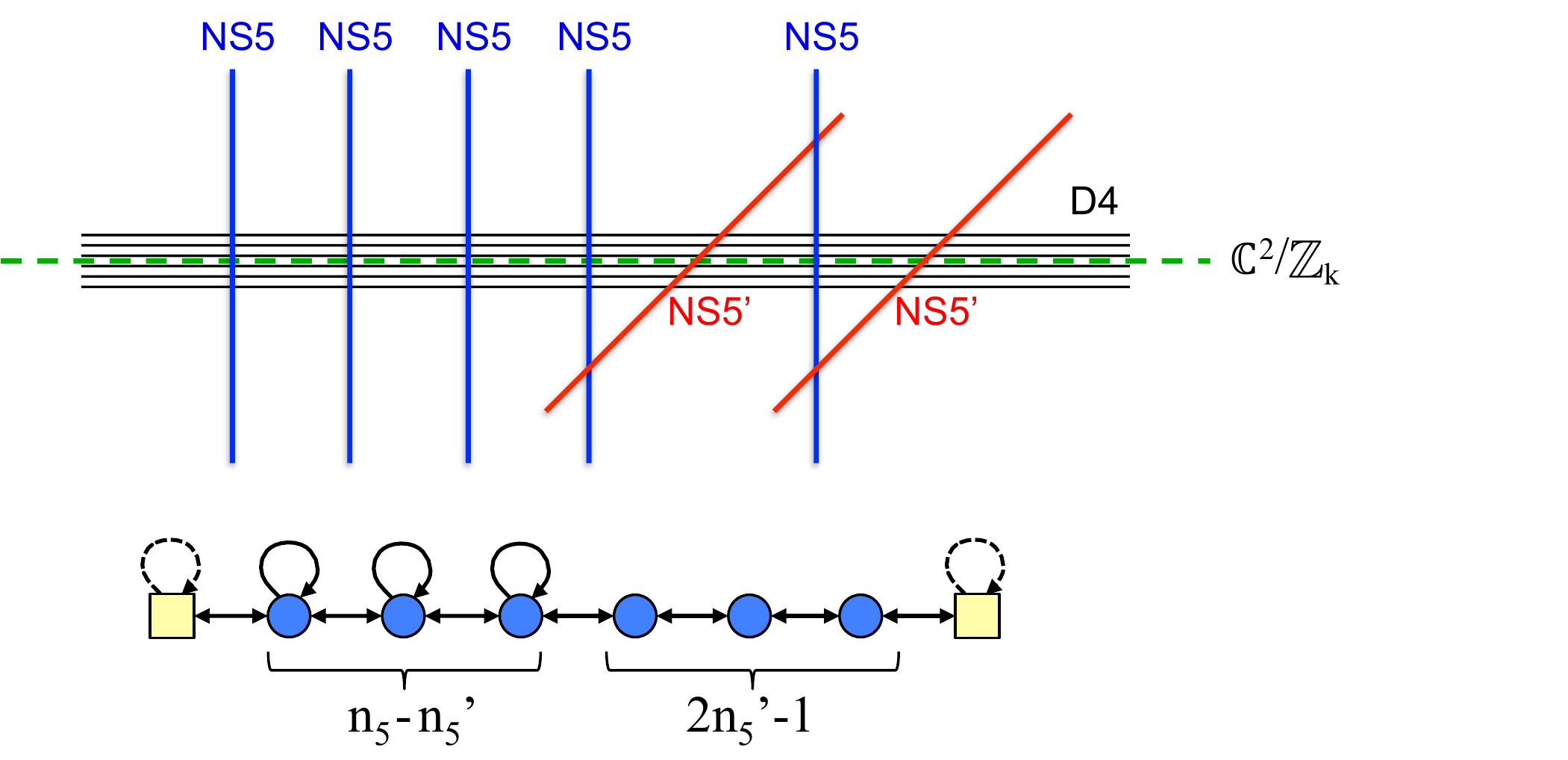}
\caption{Top: Type IIA brane configuration corresponding to a core theory with $n_5=5$ and $n_5'=2$ in the second standard ordering. Bottom: the associated linear quiver {\it before} the $\mathbb{Z}_k$ orbifold.}
\label{IIA_quiver_ordering_2}
\end{center}
\end{figure}

A second canonical ordering, shown in \fref{IIA_quiver_ordering_2}, corresponds to pairing up the $n_5'$ NS'-branes with $n_5'$ of the NS-branes and alternate them on the right hand side, leaving the remaining $n_5-n_5'$ NS-branes in the left hand side. Before the orbifold, the configuration contains an $\NN=2$ sector (vector-like bifundamentals coupled to adjoints by cubic superpotentials), coupled to a linear set of $\NN=1$ sectors (no adjoint, and quartic quartic superpotential among the vector-like bifundamentals). The $\NN=2$ sector corresponds to $n_5-n_5'$ Type II nodes and the $\NN=1$ sector consists of $2n_5'-1$ Type I nodes.

\begin{figure}[h]
\begin{center}
\includegraphics[width=15cm]{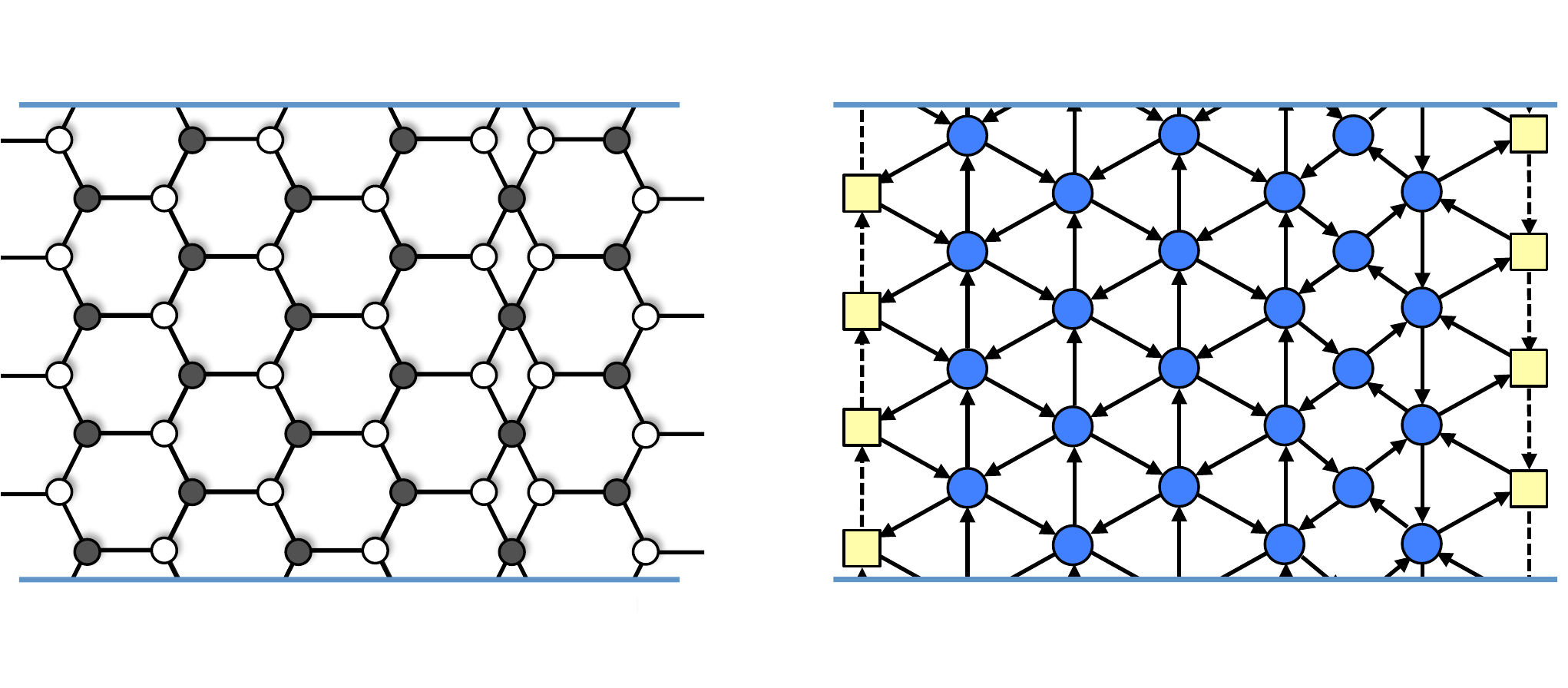}
\caption{Dimer and BFT for a core theory with $n_5=5$ and $n_5'=2$ in the first standard ordering. Every plaquette in the quiver corresponds to a superpotential term.}
\label{dimer_quiver_ordering_1}
\end{center}
\end{figure}

\begin{figure}[h]
\begin{center}
\includegraphics[width=14cm]{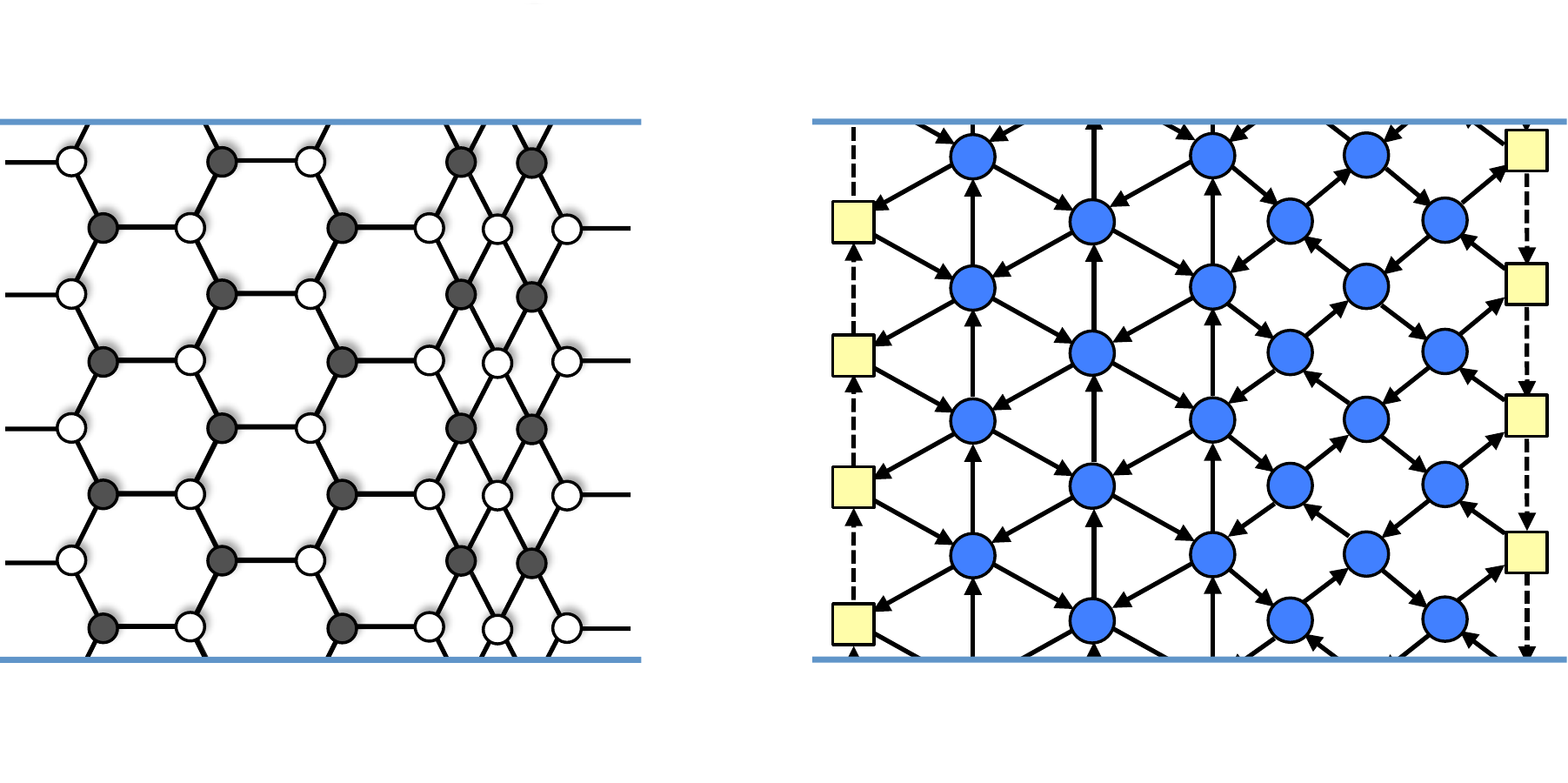}
\caption{Dimer and BFT for a core theory with $n_5=5$ and $n_5'=2$ in the second standard ordering. Every plaquette in the quiver corresponds to a superpotential term.}
\label{dimer_quiver_ordering_2}
\end{center}
\end{figure}

The two different orderings produce seemingly different parent field theories, which in turn produce seemingly different orbifold field theories. Examples of orbifold theories corresponding to the previous figures are displayed in \fref{dimer_quiver_ordering_1}, \fref{dimer_quiver_ordering_2}, for the left-right and alternating orderings, respectively.

It is obvious that the above two orderings, and in fact any other ordering, can be related by operations that exchange the position of adjacent NS- and NS'-branes. This kind of brane motion has been extensively exploited in the brane realization of supersymmetric gauge theories, and related to Seiberg duality, starting from \cite{Elitzur:1997fh}. We can therefore anticipate that the different orderings of the brane configuration correspond merely to different Seiberg dual UV descriptions of a unique underlying $\NN=1$ SCFT. This intuition will be proven explicitly in the next section.

Finally, note that to show that the ordering of punctures in the Riemann surface of the parent theory is irrelevant, one also needs to invoke operators exchanging punctures of the same kind (i.e. two members of a pair of NS-branes, or of NS'-branes). These kind of operations are exactly identical to those in \cite{Gaiotto:2015usa}, so we will not discuss them further.

\bigskip

\subsection{Effect of Seiberg Duality}
\label{sec:seiberg}

In the previous section, we have presented two explicit Seiberg dual versions of our theories. It is interesting to discuss the effect of Seiberg duality in more generality. To do so, let us start by considering the unorbifolded theories. In this case, Seiberg duality acts on Type I nodes. 

Considering all nodes in the quiver have ranks equal to $N$, Type I nodes have $N_f=2N_c$ and hence the rank remains equal to $N$ after the duality. When dualizing, we replace the electric flavors by the magnetic ones: $\tilde{X}_{i-1,i}$, $\tilde{X}_{i,i-1}$, $\tilde{X}_{i,i+1}$, $\tilde{X}_{i+1,i}$. In quiver language, this is achieved by reversing the direction of the arrows connected to the dualized node. While the unorbifolded quiver is unaffected by this operation, it becomes non-trivial for the general theories we consider.

\begin{figure}[h]
\begin{center}
\includegraphics[width=9cm]{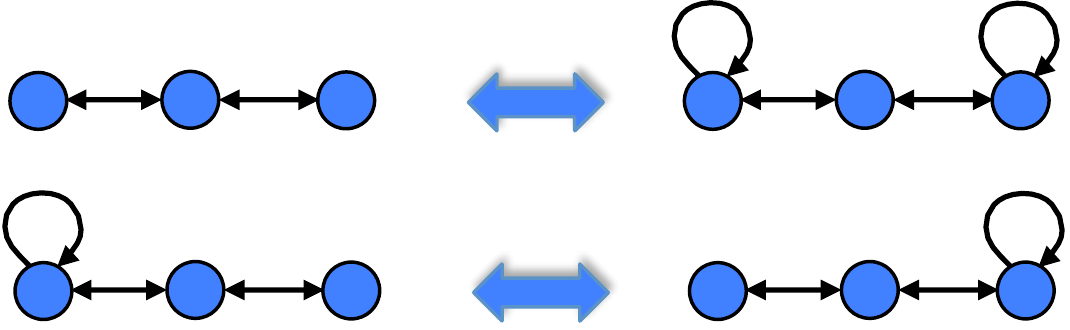}
\caption{The two possible behaviors when Seiberg dualizing a Type II node (the central nodes in the figure) in a linear quiver.}
\label{basic_Seiberg_dualities}
\end{center}
\end{figure}

We also need to incorporate mesons, which in terms of the electric flavors are given by
\beq
\begin{array}{cclcccl}
M_{i-1,i+1} & = & X_{i-1,i} X_{i,i+1} & \ \ \ \ \ \ \ & \Phi_{i-1} & = & X_{i-1,i} X_{i,i-1} \\
M_{i+1,i-1} & = & X_{i+1,i} X_{i,i-1} & \ \ \ \ \ \ \ & \Phi_{i+1} & = & X_{i+1,i} X_{i,i+1} 
\label{mesons}
\end{array}
\eeq
$M_{i-1,i+1}$ and $M_{i+1,i-1}$ are bifundamental, while $\Phi_{i-1}$ and $\Phi_{i+1}$ transform in the adjoint representations of nodes $i-1$ and $i+1$, respectively. Cubic superpotential couplings between the mesons and the magnetic flavors must be added to the superpotential. Up to signs, the new terms in the superpotential are
\beq
\Delta W= M_{i-1,i+1} \tilde{X}_{i+1,i} \tilde{X}_{i,i-1} + M_{i+1,i-1} \tilde{X}_{i-1,i} \tilde{X}_{i,i+1} + \Phi_{i-1} \tilde{X}_{i-1,i} \tilde{X}_{i,i-1} + \Phi_{i+1} \tilde{X}_{i+1,i} \tilde{X}_{i+1,i}.
\eeq
The original superpotential \eref{superI} becomes a mass term for $M_{i-1,i+1}$ and $M_{i+1,i-1}$, so they can be integrated out at low energies. The fate of the adjoint mesons $\Phi_{i-1}$ and $\Phi_{i+1}$ depends on whether the nearest neighbor nodes are Type I or II. It is straightforward to see that the net effect is to simple change the type of each of these nodes, leading to the two possibilities illustrated in \fref{basic_Seiberg_dualities}. If the dualized node is surrounded by two Type I/Type II nodes, they are turned into Type II/Type I nodes. The number of nodes of each type is in generally not preserved, as clearly shown by the explicit examples in \sref{sec:two-orderings}. When the two nearest neighbors are nodes of different types, the effect of Seiberg duality is to switch them. We can use this process to move nodes of different types along the linear quiver, without changing the number of nodes of each type.

The previous discussion is nicely translated into the Type IIA brane configuration. As previously explained, Type I nodes correspond to stacks of D4-branes stretched between a pair of rotated NS5-branes. Seiberg duality corresponds to exchanging the position of both NS5-branes \cite{Elitzur:1997fh}, which in turn naturally results in the change of type of the adjacent nodes.

Let us now consider the orbifolded theory, in which each node of the original linear quiver gives rise to a column of $k$ nodes. Our previous analysis straightforwardly extends to dualizations of all nodes in a given column of Type I. We thus see that dualizations of such columns simply change the types of the two adjacent ones. The analysis can also be carried out graphically using quiver/dimer technology, by applying the tools in \cite{Beasley:2001zp,Feng:2001bn,Franco:2005rj}.

It is interesting to notice that more general patterns of dualizations of nodes of Type I are possible, in particular sequences that do not involve dualizing entire columns. These operations do not have a counterpart in the unorbifolded Type IIA brane configuration. In addition, it is now also possible to Seiberg dualize type II nodes, since they no longer involve adjoint chiral fields. Doing so is straightforward and very interesting. It leads to theories that are not described by bipartite graphs on a cylinder and we will not discuss this possibility any further in this article.

\bigskip

\section{Global Symmetries, Punctures and Zig-Zag Paths}

\label{sec:global-symm}

A crucial property of SCFTs is their structure of global symmetries. They can be determined using the UV description in some weak coupling regime. Global symmetries free of mixed anomalies with the gauge groups are also an important ingredient in the computation of the superconformal index, which we will undertake in section \sref{sec:index}. In this section, we study in detail the global symmetries of core class $\mathcal{S}^1_k$ theories, although several results hold beyond them.

First of all, these theories contain an $SU(N)^{2k}$ global symmetry that is manifestly realized in the quiver as non-dynamical gauge nodes, represented by squares in the quiver diagrams and external faces in the dimer. In addition, the theories contain abelian global symmetries. Remarkably, the fact that the theories we are studying are BFTs enables a combinatorial determination of their global symmetries. In what follows, we introduce a useful prescription for identify anomaly-free abelian global symmetries of the class of ${\cal S}^1_k$ field theories (in particular including those in \cite{Gaiotto:2015usa}\footnote{In fact, related ideas have been already used in \cite{Gaiotto:2015usa}. More generally, similar ideas have been applied for a long time in the context of certain classes of BFTs, since \cite{Butti:2005vn}.}). This information also provides clear definitions, at least for the core theories, of the minimal and maximal punctures in the associated Riemann surface.\footnote{There are alternative realizations of minimal punctures based on closing non-minimal punctures by higgsing as in \cite{Gaiotto:2015usa}.}

For any bipartite graph on a Riemann surface, it is possible to define a set of {\it zig-zag paths}. These are oriented paths defined by the property that they cross edges in their middle point, and turn maximally to the left on white nodes and maximally to the right on black nodes. As a result of this definition:
\begin{itemize}
\item Every edge is crossed by a two zig-zag paths, running in opposite directions. 
\item Nodes in the graph are contained inside disks on the Riemann surface whose boundaries are made out of zig-zag paths and are oriented clockwise and counterclockwise for white and black nodes, respectively.
\end{itemize}
As an aside, it is interesting to remark that in various contexts, such as brane tilings \cite{Hanany:2005ss,Feng:2005gw} and Postnikov diagrams for the positive Grassmannian \cite{2006math......9764P}, these two properties have been exploited for reconstructing the underlying bipartite graphs starting from zig-zag paths. It would be interesting to explore whether such ideas have useful applications for class $\mathcal{S}^1_k$ theories.

\begin{figure}[h]
\begin{center}
\includegraphics[height=7.5cm]{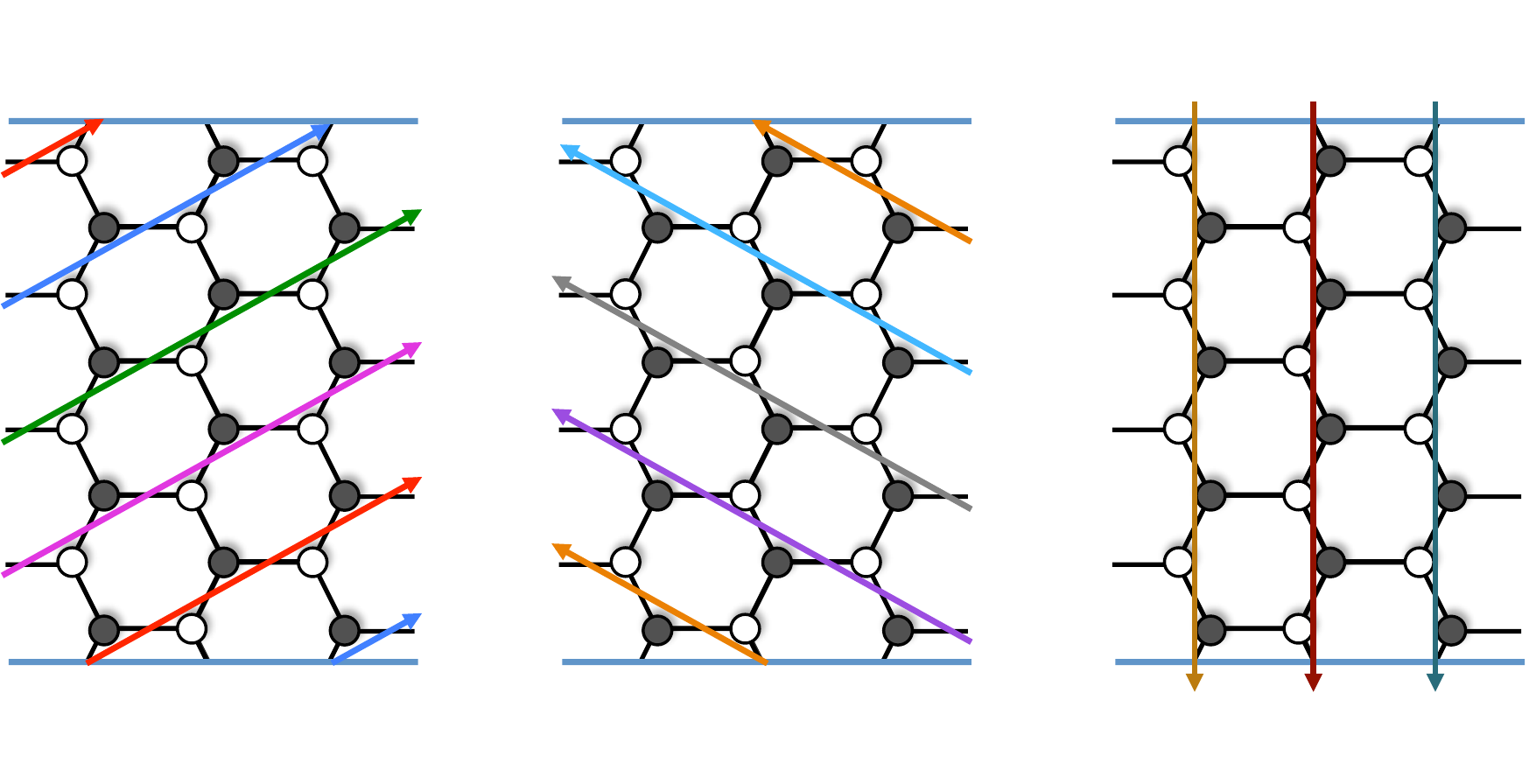}
\vspace{-.5cm}\caption{The zig-zag paths for one of the theories in \cite{Gaiotto:2015usa} with $k=4$. The paths in the first two pictures are associated to the intrinsic symmetries (of the underlying 6d theory), and those in the third figure correspond to minimal punctures.}
\label{sk-paths}
\end{center}
\end{figure}

Zig-zag paths can be of two kinds: closed (defining a non-trivial homology cycle on the Riemann surface tiled by the bipartite graph) or open (i.e. with endpoints on external faces).\footnote{Formally, zig-zag paths define non-trivial classes in the relative homology group $H_1(\Sigma;L)$ of the surface $\Sigma$ with $L$ corresponding to 1-cycles defined by sequences of non-dynamical nodes.} In \fref{sk-paths} we show the  zig-zag paths for a theory of the kind considered in \cite{Gaiotto:2015usa}, whose global symmetry structure is easily recovered by applying our discussion below to these paths. 

\begin{figure}[h]
\begin{center}
\includegraphics[height=6.5cm]{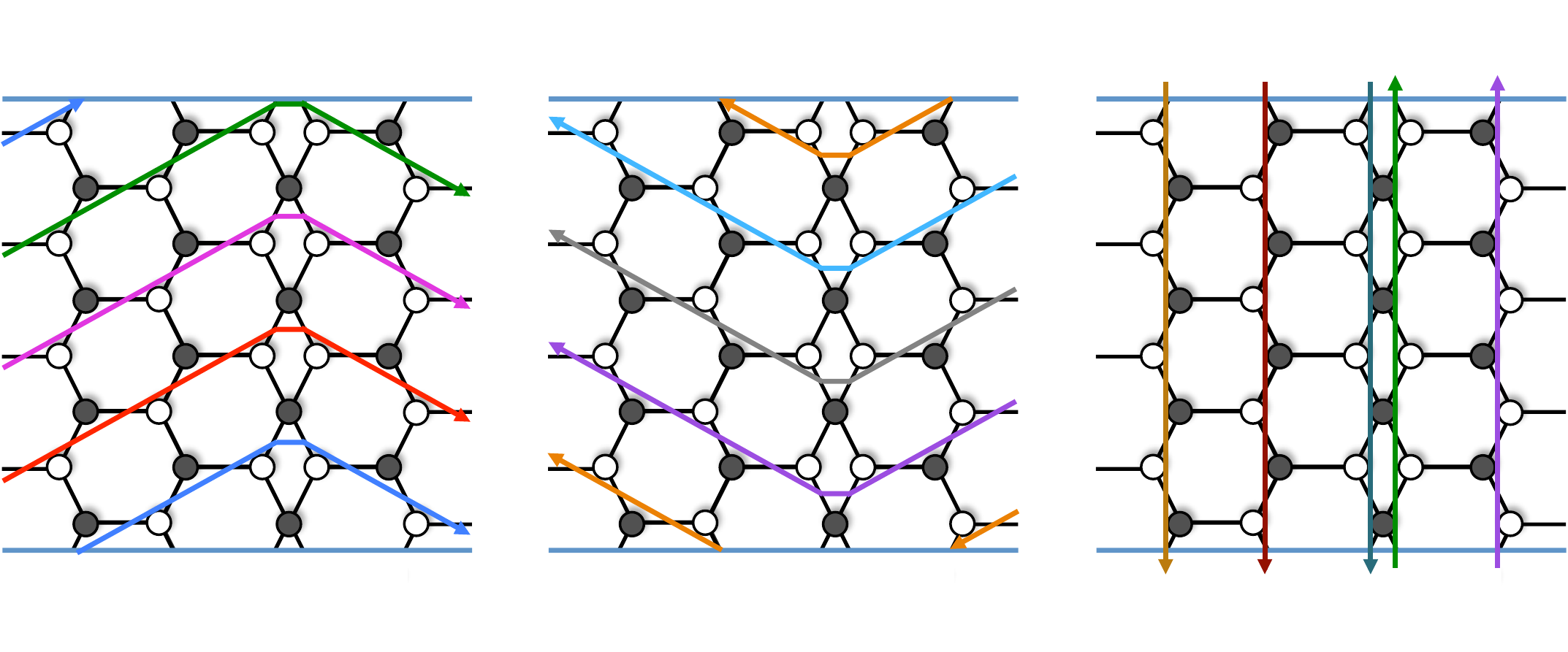}
\vspace{-1cm}\caption{The zig-zag paths for a class ${\cal S}^1_k$ theory with $k=4$, corresponding to a configuration with $n_5=3$ and $n_5'=2$ in the first canonical ordering. The paths in the first two pictures are associated to intrinsic symmetries, and those in the third figure correspond to minimal punctures (of two kinds, distinguished by the path orientation).}
\label{example2-paths}
\end{center}
\end{figure}

\begin{figure}[h]
\begin{center}
\includegraphics[height=7.5cm]{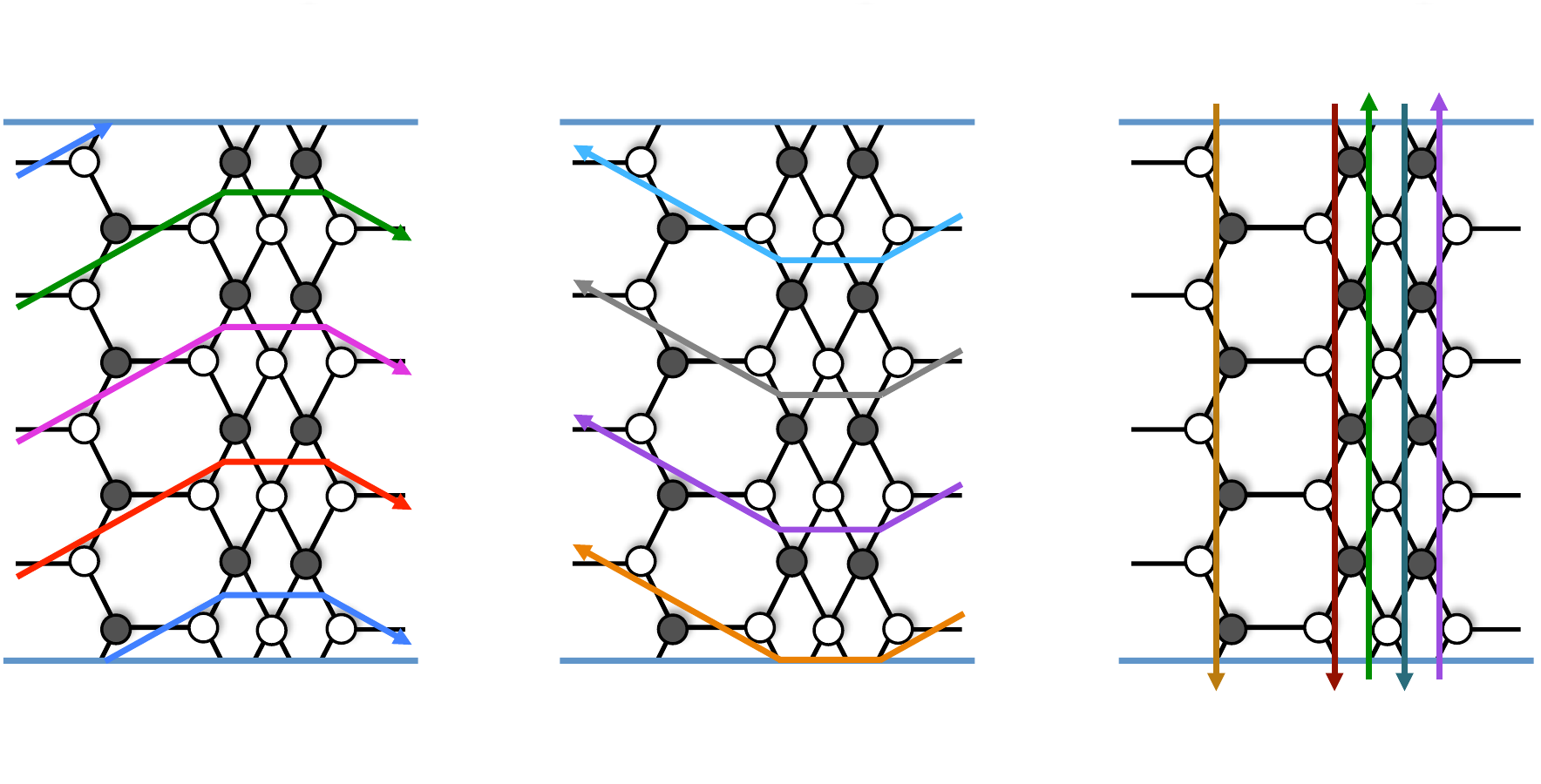}
\vspace{-.5cm}\caption{The zig-zag paths for a class ${\cal S}^1_k$ theory with $k=4$, corresponding to a configuration with $n_5=3$ and $n_5'=2$ in the second canonical ordering. The paths in the first two pictures are associated to intrinsic symmetries, and those in the third figure correspond to minimal punctures of two kinds.}
\label{example1-paths}
\end{center}
\end{figure}

In Figures \ref{example2-paths} and \ref{example1-paths} we display the zig-zag paths for two examples of our more general ${\cal S}^1_k$ theories. In fact, they both follow from Type IIA configurations with $n_5=3$ and $n_5'=2$, but correspond to the two different canonical orderings we discussed earlier. Notice that the endpoints of open zig-zag paths are identical in both theories, i.e. they connect the same pairs of global symmetry nodes, although they differ in their internal trajectories. Furthermore, the two theories have the same pattern of closed paths, modulo reordering. As we discuss below, this implies that the two theories have identical global symmetry structures. This is expected since it is ultimately a manifestation of the fact that the two theories are related by reshuffling NS5-branes in the Type IIA setup or, as explained in section \ref{sec:seiberg}, that they are different Seiberg dual phases of the same underlying SCFT. Furthermore, this is also related to our earlier comments regarding the reconstruction of bipartite graphs in terms of zig-zag paths. Seiberg dualities manifest as reorganizations and deformations of zig-zag paths inside the bulk of the graph.

Every zig-zag path defines an anomaly free $U(1)$ global symmetry as follows. The charge of a bifundamental chiral multiplet under the $U(1)$ symmetry associated to a path is given by the intersection number of the corresponding edge and the zig-zag path, counted with orientation. In other words, if the edge is not crossed by the path, the charge is zero; if it is crossed by the path, the charge is $\pm 1$ depending on the relative orientation between the edge and the path (e.g. adopting the convention that edges are oriented from white to black nodes). It is clear that the sum of the $U(1)$ generators associated to all zig-zag paths in a graph is equal to zero. Thus, one of the corresponding $U(1)$ symmetries is redundant.
We have now all the tools that are necessary for discussing the global $U(1)$ symmetries of the core ${\cal S}^1_k$ theories. From the examples in Figures \ref{example2-paths} and \ref{example1-paths}, it is clear that a general theory has the following zig-zag paths:

\begin{itemize}
\item $k$ open paths going from left to right.
\item $k$ open paths going from right to left.
\item $n+1$ vertical closed paths ($n_5$ and $n_5'$ in each direction).
\end{itemize}

Recalling that one of these symmetries is redundant, we conclude a general theory has a $U(1)^{2k+n}$ global symmetry. This symmetry is in precise agreement with the 6d description, from which we expect a $[U(1)^k/U(1)]\times [U(1)^k/U(1)] \times U(1)_t$ Cartan subgroup intrinsic symmetry of the ${\cal T}^N_k$ theory and a $U(1)$ for each minimal puncture, resulting in an additional $U(1)^{n+1}$. Interesting, for core theories we can establish a precise map between the topology of zig-zag paths and different types of global symmetries, as follows:

\begin{itemize}

\item Closed zig-zag paths correspond to global symmetries associated to simple punctures of the Riemann surface. In the Type IIA configuration, in weak coupling regimes, these are associated to the position of NS5-branes. Simple punctures corresponding to NS- and NS'-branes are associated with paths with opposite orientations. On the cylinder, the closed paths are vertical, and are associated to columns of gauge factors descending from a node of the parent theory before orbifolding. There is an associated anomaly-free global $U(1)$ which corresponds to the combination of the $U(1)$'s in the $U(N)$ factors of the corresponding column, arising in the D-branes realization, c.f. footnotes \ref{note-u1} and \ref{note-u1k}.

\item Each of the $[U(1)^k/U(1)]$ intrinsic symmetries can be identified with the $k$ left-right or right-left open zig-zag paths divided by their diagonal combination. Finally, $U(1)_t$ is given by the anti-diagonal combination of left-right and right-left paths.

\end{itemize}
The previous discussion is basically identical to the one presented in \cite{Gaiotto:2015usa} for theories and zig-zag paths of the kind shown in \fref{sk-paths}. 

An important feature of zig-zag paths that might turn out to be useful in future applications is that they allow the identification of global symmetries in BFTS defined in terms of arbitrary Riemann surfaces.

\bigskip

\section{Marginal Deformations}
\label{sec:marginal}

One of the remarkable features of the linear $\NN=2$ theories, which also extends to the $\NN=1$ linear quiver theories, is that many of their properties of the SCFTs can be encoded in a punctured sphere $\Sigma$, with 2 maximal punctures and $(n+1)$ minimal ones). For instance, the number of marginal couplings is given by the number of complex structure parameters of $\Sigma$, namely $n$. Moreover, the degeneration limits of the Riemann surface correspond to regimes where a gauge factor becomes weakly coupled, with gauge coupling controlled by the corresponding marginal coupling.

These features descend to the orbifold theories, which are associated to a sphere $\Sigma$ with 2 maximal punctures and $n_5+n_5'$ minimal ones. Namely, the complex structure parameters of $\Sigma$ should match the field theory exactly marginal couplings.  To compute the latter, we note that the theories in general have non-trivial anomalous dimensions, differing from free field values, requiring the use of the exact NSVZ beta functions. Actually, it is straightforward to count the number of exactly marginal deformations of the orbifold theory by the method in \cite{Leigh:1995ep} (for a similar analysis predating that in \cite{Gaiotto:2015usa}, see \cite{Hanany:1998ru}), as follows. 

First note that we have $nk$ gauge couplings and $(2n-\tilde{n})k$ superpotential couplings, which gives a parameter space of complex dimension $(3n - \tilde{n})k$. We then impose that all the beta functions for the gauge couplings and the superpotential couplings vanish. The vanishing of the beta functions for the Type I nodes implies
\begin{equation}
A_{g_{I, a}} = N + \frac{1}{2}\sum \gamma_{I. a} = 0,
\end{equation}
where the sum is taken over the $4N$ chiral multiplets coupled to a Type I node, and $\gamma$ represents the anomalous dimension of a field. $a$ labels the gauge nodes within a given column.  On the other hand, the vanishing of the beta functions for the Type II nodes becomes
\begin{equation}
A_{g_{II, a}} = \frac{1}{2}\sum \gamma_{II. a} = 0,
\end{equation}
where the sum is taken over the $6N$ chiral multiplets coupled to a Type II node. We further impose that the beta functions for the quartic superpotential \eqref{superI} vanish
\begin{equation}
A_{\lambda_{I, a}} = 1 + \frac{1}{2}\sum \gamma_{I, a} = 0,
\end{equation}
where the sum is taken over the four chiral multiplets which make the quartic superpotential \eqref{superI}. The vanishing of the beta functions for the cubic superpotential \eqref{superII} is 
\begin{eqnarray}
A_{\lambda_{II\text{-}1, a}} &=& \frac{1}{2}\sum \gamma_{II\text{-}1, a} = 0,\\
A_{\lambda_{II\text{-}2, a}} &=& \frac{1}{2}\sum \gamma_{II\text{-}2, a} = 0,
\end{eqnarray}
where the first and the second conditions represent the vanishing of the beta function for the first and the second terms in \eqref{superII} respectively. The sum is taken over the three chiral multiplets in \eqref{superII} correspondingly. In total, we have $(3n - \tilde{n})$ conditions. 

However, not all of them are independent. In fact, as can be easily checked from \fref{orbifold-nodes}, for each column of gauge nodes, we have a condition
\begin{equation}
\sum_{a=1}^k\frac{A_{g_{I,a}}}{N} = \sum_{a=1}^k A_{\lambda_{I, a}},
\end{equation}  
or
\begin{equation}
\sum_{a=1}^k\frac{A_{g_{II,a}}}{N} = \sum_{a=1}^k\left(A_{\lambda_{II\text{-}1, a}} +A_{\lambda_{II\text{-}2, a}}\right).
\end{equation}
Therefore, among the $(3n - \tilde{n})k$ conditions, $n$ of them are redundant. One can also see that there are same number of phase rotations that can be removed. Hence, we have in total $(3n - \tilde{n})k - n$ complex conditions. Putting everything together, the complex dimension of the conformal manifold is $n = n_5 + n_5'- 1$. This agrees with the number of the complex structure moduli of a sphere with $n_5 + n_5'$ simple punctures and two maximal punctures. 

The correspondence between marginal couplings (which are associated to columns of gauge nodes) and punctures is manifest because the latter are defined in terms of vertical zig-zag paths, as explained in \sref{sec:global-symm}, so the $(n+1)$ simple punctures define $n$ columns of gauge factors. Therefore, the marginal couplings associated to the punctures can be used to define weak coupling limits of particular columns of gauge factors.

\bigskip

\section{Constructing Theories from Basic Building Blocks}

\label{section_theories_from_building_blocks}

In this section we study how to generate new theories by either gluing maximal punctures or closing minimal ones.

\bigskip

\subsection{Free Trinion}

\label{free_trinion}

\begin{figure}[h]
\begin{center}
\includegraphics[width=6cm]{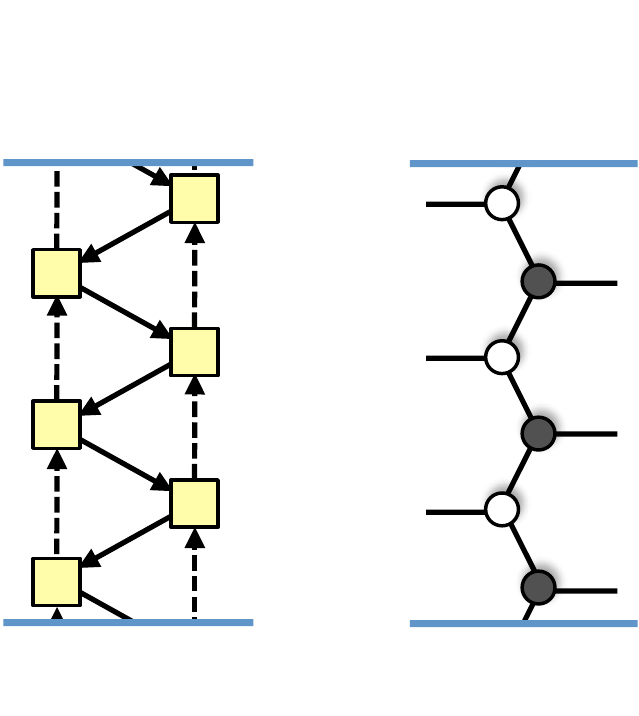}
\vspace{-.5cm}\caption{The free trinion for $k=3$.}
\label{trinionk3}
\end{center}
\end{figure}

The basic building block for constructing a large class of $\mathcal{S}_k$ theories in \cite{Gaiotto:2015usa} is the free trinion, which we show in \fref{trinionk3}, a configuration describing an $\NN=2$ free theory which can be glued to build interacting theories. It is given by a sphere with two maximal and one minimal puncture, and is related to a Type IIA configuration of one NS5-brane with two stacks of $N$ semi-infinite D4-branes sticking out from its sides. The process of gluing trinions in $\NN=2$ theories amounts to bringing in more NS5-branes from infinity in order to bound finite stacks of D4-branes among them. Alternatively, it corresponds to gauging diagonal combinations of global symmetry factors in different trinions. Hence, although the trinion describes a free theory of one $\NN=2$ hypermultiplet (or its $\IZ_k$ orbifold), by gluing several trinions one can construct non-trivial interacting SCFTs.

In our core class ${\cal S}^1_k$ theories, we have two kinds of NS5-branes, denoted NS- and NS'-branes, which  correspond to two kinds of minimal punctures in the Riemann surfaces. One is tempted to conclude that these theories therefore require two (or more) kinds of trinions to construct them, depending on the kind of minimal puncture they introduce. We will show that the whole class of theories can be built with only one kind of trinion, which is exactly the $\NN=2$ trinion, but with two inequivalent gluing prescriptions, which we will call $\NN=2$ and $\NN=1$ gluing. This will be discussed in more detail in \sref{sec:gluing-maximal}. This viewpoint agrees with the Type IIA brane configuration picture, where the local configuration near an NS- and an NS'-brane are isomorphic, and it is only through the choice of how to glue different configurations to form new finite intervals that either $\NN=1$ or $\NN=2$ sectors arise. Other formulation with several basic trinions may be possible but are equivalent to ours\footnote{In fact, a simple modification is to different trinions differing in whether they include or not the information about the mesonic bi-fundamental fields corresponding to vertical arrows. We choose to introduce just only one trinion and include the bi-fundamentals when needed, namely upon $\NN=2$ gluing or as an extra dressing of maximal punctures (describing  e.g. Type IIA systems of D4-branes suspended between NS'- and D6-branes).}.

\bigskip

\subsection{Gluing Maximal Punctures}
\label{sec:gluing-maximal}

Let us now consider the gluing of maximal punctures in core theories. This process was discuss in \cite{Gaiotto:2015usa} for a class of $\mathcal{S}_k$ theories. Not surprisingly, since the ${\mathcal{S}}^1_k$ class generalizes them, an additional possibility for gluing arises. %

It is instructive to first consider the theories before the $\mathbb{Z}_k$ orbifold. In this case, there are two qualitatively different ways of gluing Type IIA brane configurations along maximal punctures, depending on the relative orientations of the NS5-branes adjacent to the glued punctures. If the two NS5-branes are parallel, the final brane configuration locally preserves $\mathcal{N}=2$ SUSY. Denoting $SU(N)$ and $\widehat{SU(N)}$ the global symmetries associated to each of the two punctures, in this case we gauge the diagonal subgroup of $SU(N)\times \widehat{SU(N)}$ using an $\mathcal{N}=2$ vector multiplet. If, instead, one of the maximal punctures to be glued is adjacent to an NS5-brane while the other one is adjacent to an NS5'-brane, the resulting configuration locally preserves only $\mathcal{N}=1$ SUSY. The diagonal subgroup of $SU(N)\times \widehat{SU(N)}$ is gauged with an $\mathcal{N}=1$ vector multiplet. This can be alternatively interpreted as gauging with a $\mathcal{N}=2$ vector multiplet and then giving an infinite mass to the $\mathcal{N}=1$ adjoint chiral multiplet it contains. These two possibilities have been introduced in \cite{Benini:2009mz} and studied in various contexts \cite{Bah:2012dg,Beem:2012yn,Xie:2013gma}. We refer to them as $\mathcal{N}=2$ and $\mathcal{N}=1$ gluing.

Below we discuss these two gluings in the presence of the $\mathbb{Z}_k$ orbifold. While for $k>1$ the theories only preserve $\mathcal{N}=1$ SUSY, we will still refer to them as $\mathcal{N}=2$ and $\mathcal{N}=1$ gluings. We will see that from the perspective of the corresponding bipartite graphs, the $\mathcal{N}=2$  and $\mathcal{N}=1$ gluings generate a column of hexagons and squares, respectively. When gluing, we will flip the node colors in one of the components whenever necessary; this is equivalent to reversing the orientation of all the arrows in the dual quiver.\footnote{The operations we discuss also apply to the gluing of the two maximal punctures in a given core theory.}

\bigskip

\subsubsection{$\mathcal{N}=2$ Gluing}

\begin{figure}[h]
\begin{center}
\includegraphics[width=14cm]{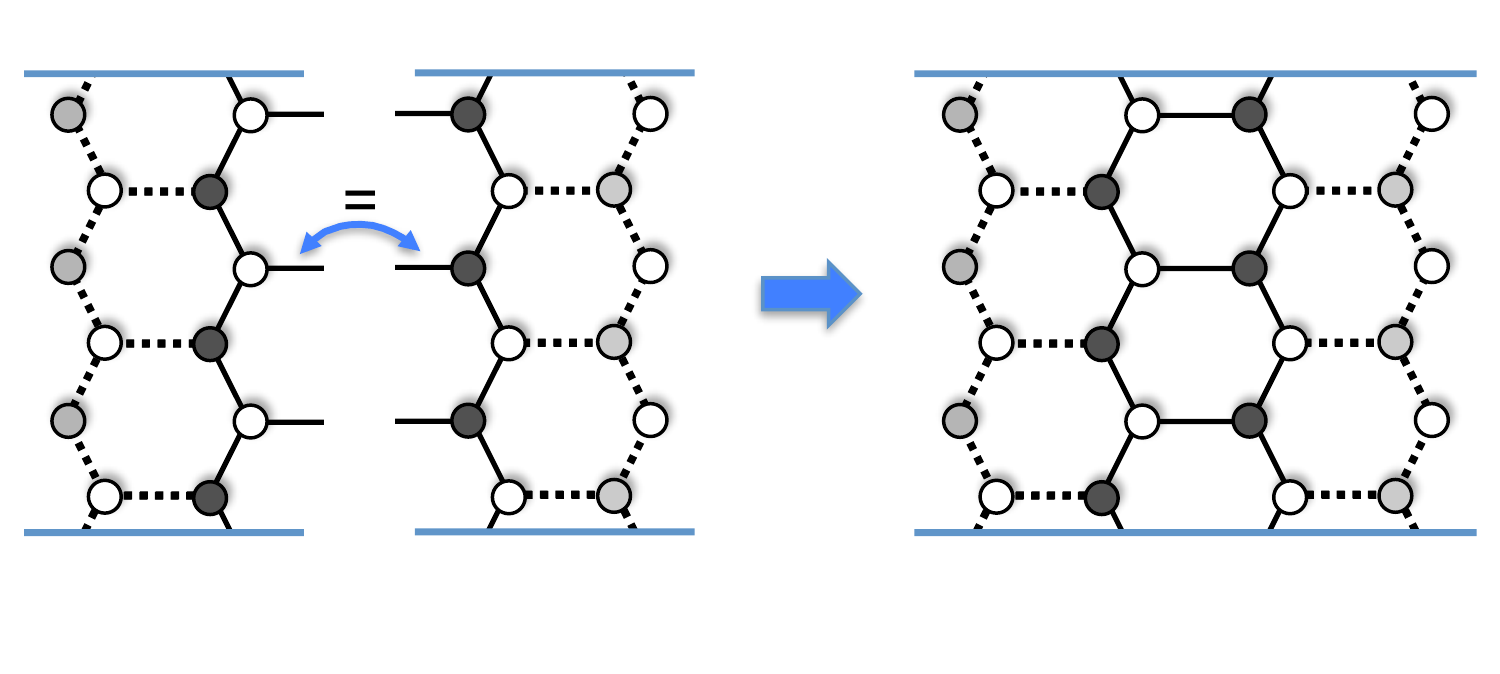}
\vspace{-1cm}\caption{$\mathcal{N}=2$ gluing. It results in a new column of hexagons.}
\label{N=2_gluing}
\end{center}
\end{figure}

Let us first discuss the $\mathcal{N}=2$ gluing, which is precisely the one discussed in \cite{Gaiotto:2015usa}. \fref{N=2_gluing} sketches its implementation in terms of the bipartite graph. The dotted lines represent the rest of the graphs. The nature of the gluing is independent of the type of faces in the columns closest to the maximal punctures to be glued. For concreteness, we show the case in which they are hexagons on both sides. Nothing in our discussion changes if these columns consist of squares in one or both sides.

The gluing is defined as follows:
\begin{itemize}
\item Make the chiral fields associated to external legs dynamical and identify them.\footnote{Notice that, as already mentioned in \sref{sec:general-class}, these fields and their superpotential couplings were not included in the theories in \cite{Gaiotto:2015usa}. As a result, they had to be explicitly incorporated when gluing maximal punctures.} These legs must be connected to nodes of different colors in each of the two theories. This ensures that the corresponding bifundamental fields have the same orientations in the two components. If this is not the case, we simply flip the color of all nodes in one of the theories. Notice that there are $k$ discrete choices for how to glue the two punctures, depending on how their external legs are paired.
\item Gauge the diagonal subgroups of global nodes pairs using $\mathcal{N}=1$ vector multiplets.
\end{itemize}
This definition is equivalent to the one introduced in \cite{Gaiotto:2015usa}. 
This process leads to a new column of hexagons, whose horizontal edges arise from the glued external legs. These edges correspond to bifundamental fields that are the $\mathbb{Z}_k$ images of the adjoint chiral field in the $\mathcal{N}=2$ vector multiplet of the unorbifolded theory.

\bigskip

\subsubsection{$\mathcal{N}=1$ Gluing}

\begin{figure}[h]
\begin{center}
\includegraphics[width=14cm]{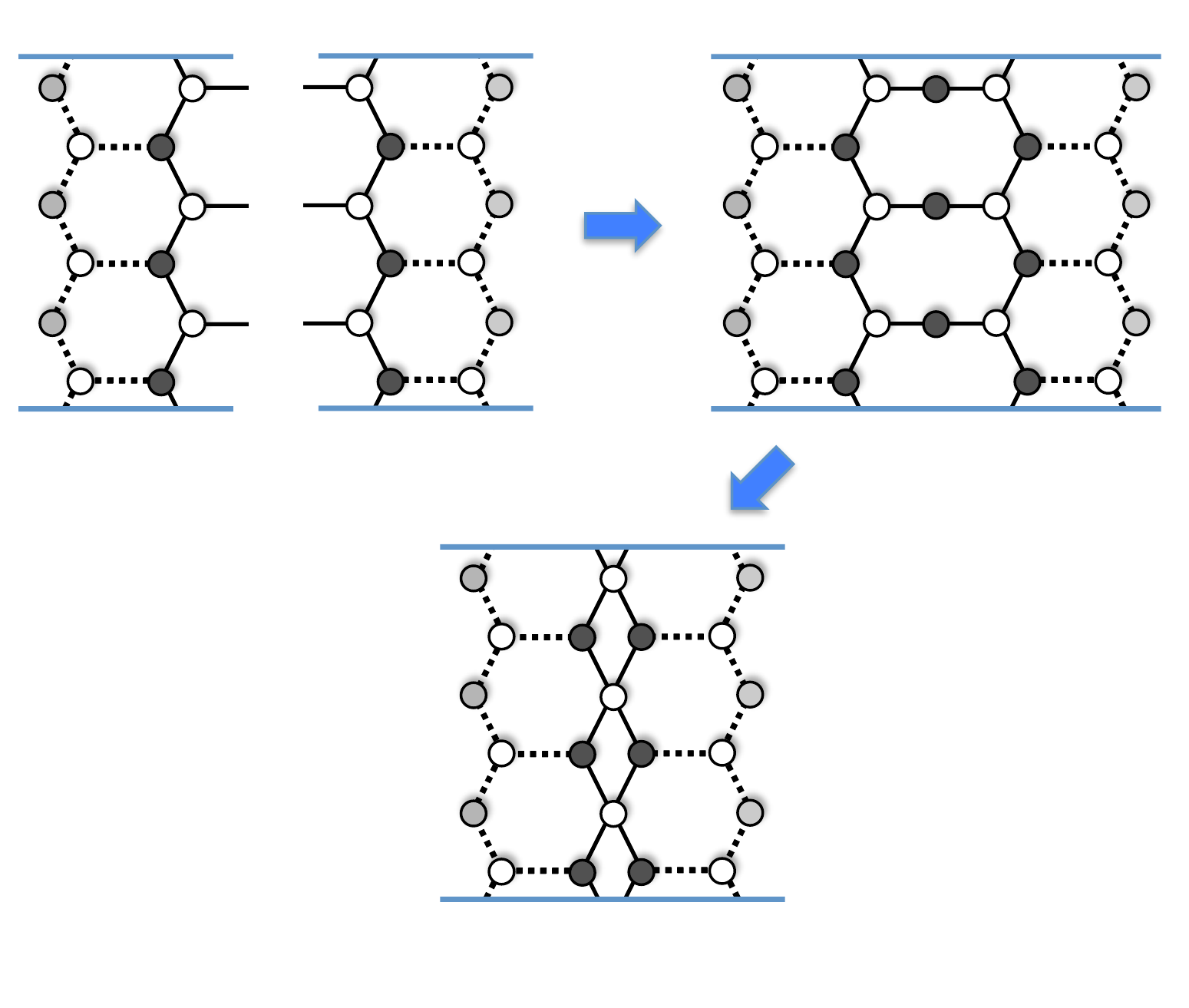}
\vspace{-1cm}\caption{$\mathcal{N}=1$ gluing. After integrating out the massive chiral fields, it results in a new column of squares.}
\label{N=1_gluing}
\end{center}
\end{figure}

The $\mathcal{N}=1$ gluing is shown in \fref{N=1_gluing}. It is defined as follows:

\begin{itemize}
\item Make the chiral fields associated to external legs dynamical and introduce quadratic (i.e. mass) terms in the superpotential coupling them, which correspond to 2-valent nodes. The legs must be initially connected to nodes of the same color in both theories. As before, if this is not the case we simply flip the color of all nodes in one of the components. Once again, there are $k$ different choices for how to glue the two punctures.
\item Gauge the diagonal subgroups of global nodes pairs using $\mathcal{N}=1$ vector multiplets.
\end{itemize}
Once again, the precise structure of the dotted pieces of the graphs is unimportant.

Since we are interested in the IR dynamics of these theories, we can integrate out the massive fields. This corresponds to removing the massive edges and merging the nodes at both sides of 2-valent nodes. We refer the reader to \cite{Franco:2012mm} for thorough discussions of this process. 
Remarkably, this operation is the $\mathbb{Z}_k$ image of giving a mass to the adjoint chiral field in the $\mathcal{N}=2$ vector multiplet of the unorbifolded theory. As shown in \fref{N=1_gluing}, all the faces in the new column are turned into squares. 

In summary, the two gluings we have discussed are the $\mathbb{Z}_k$ orbifolded versions of the $\mathcal{N}=2$ and $\mathcal{N}=1$ gluings of the parent theories. It is hence natural to interpret them as {\it deconstructions} of the gluings in the unorbifolded theories along the lines of \cite{ArkaniHamed:2001ca,ArkaniHamed:2001ie}. 

\bigskip

\subsubsection{Gluing Beyond Core Theories.}

Although we have defined the two types of gluing in terms of core theories, it is natural to extend them to more general quiver gauge theories. In particular, we will promote the definitions above in the obvious way to the gluing of any two theories containing a periodic array of $k$ $SU(N)$ global symmetry nodes, each of them having two non-dynamical arrows as in core theories. Notice that this definition does not say anything about the number of dynamical arrows connected to the global nodes. In terms of bipartite graphs, this implies that we can glue external faces that have different shapes from those appearing in core theories.

\bigskip

\subsection{Closing Minimal Punctures by Higgsing}

\label{section_closing_minimal_punctures}

In this section we discuss the closing of minimal punctures. It is natural to consider closing of minimal punctures that keep the theories within the core ${\cal S}_k^1$ class. Given the connection with the Type IIA brane configuration, it is clear that such puncture closings corresponds in the gauge theory to an orbifold-invariant higgsing, namely one that involves vevs for $k$ edges distributed along the periodic direction of the dimer/quiver diagram. In our discussion, such higgsings are always of baryonic kind, i.e. they are triggered by (single or multiple) vevs for the dibaryons built out of the bifundamental field associated to the corresponding edge.  Depending on the kind of columns this series of edges separates (hexagons or squares) we can have three variants of such higgsings. In the presence of additional punctures of either kind, one can use Seiberg dualities to permute and change the type of the punctures, as discussed in \sref{sec:seiberg}, and all the different higgsings turn out to be related; however, we prefer to keep the discussion explicit and discuss them independently. 

\begin{figure}[h]
\begin{center}
\includegraphics[height=5.5cm]{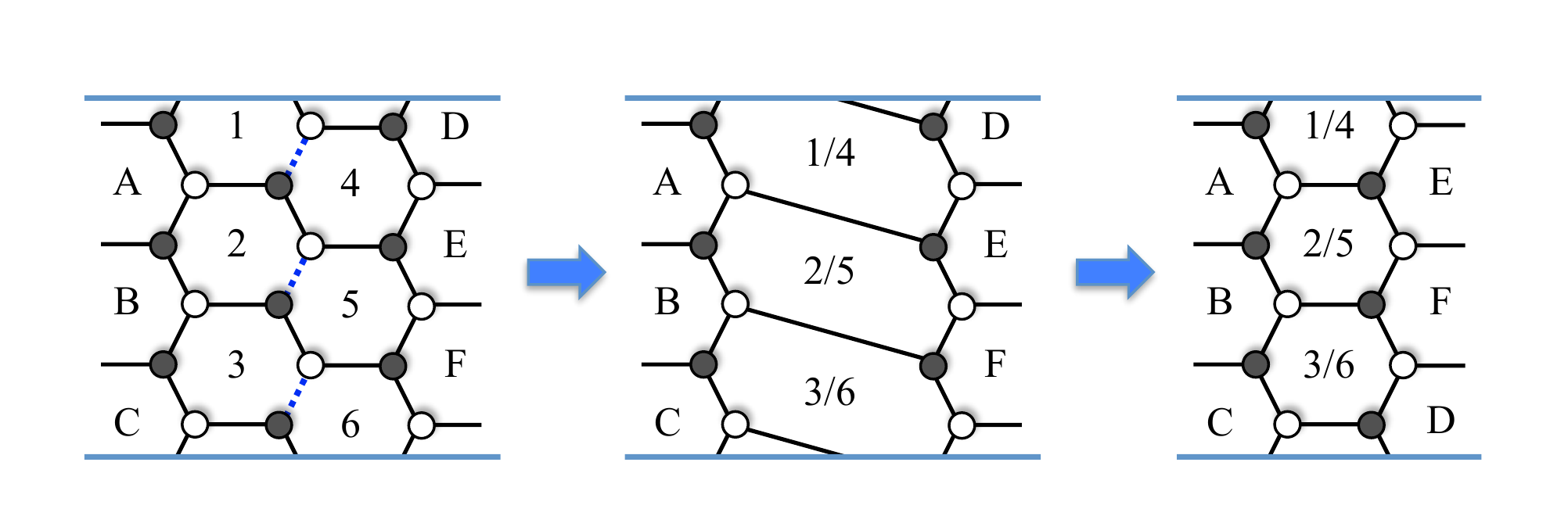}
\vspace{-.7cm}\caption{A higgsing in the dimer corresponding to removal of a puncture separating two sectors with $\NN=2$ gluing. Invariance under the $\IZ_k$ cyclic symmetry of the theory implies the result still falls in a core theory.}
\label{hh}
\end{center}
\end{figure}

Consider first the case of a puncture separating two columns of hexagons, as shown in \fref{hh}. The Higgs mechanism combining the associated punctures corresponds to giving vevs to alternating edges, which we show as dotted blue lines in the figure, in the vertical series of edges separating the hexagon columns. This results in a set of nodes becoming 2-valent, which map to mass terms in the superpotential. Integrating out the massive fields corresponds to condensing the nodes at the endpoints of the 2-valent nodes. The final graph corresponds to fusing adjacent pairs of hexagons into single hexagons. The resulting theory contains all punctures of the original one except for the one we wished to remove. 

\begin{figure}[h]
\begin{center}
\includegraphics[height=5.5cm]{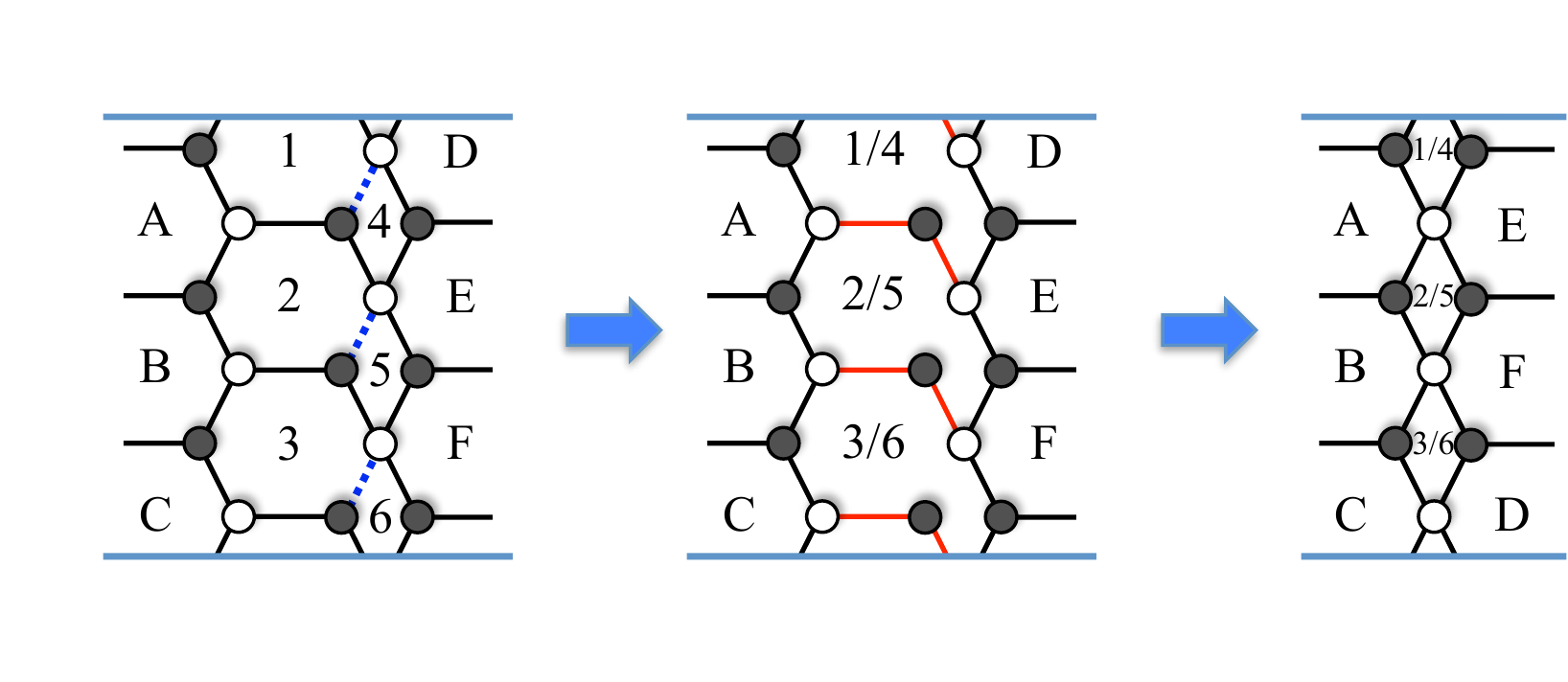}
\vspace{-.7cm}\caption{Higgsing in the dimer corresponding to removal of a puncture separating sectors with $\NN=2$ and $\NN=1$ gluings. For clarity, we include the intermediate step at which massive fields, shown in red, have not been integrated out yet. The final result is a core theory, since the higgsing is invariant under the $\IZ_k$ cyclic symmetry.}
\label{hs}
\end{center}
\end{figure}

Consider the case of a puncture separating one column of hexagons from a column of squares, see \fref{hs}. The Higgs mechanism by vevs of alternating edges in the vertical series makes the nodes 2-valent. Integrating out the massive fields contracts the diagram by fusing each hexagon with its adjacent square into a single square. The resulting theory is such that the puncture has been closed off.

\begin{figure}[h]
\begin{center}
\includegraphics[height=5.5cm]{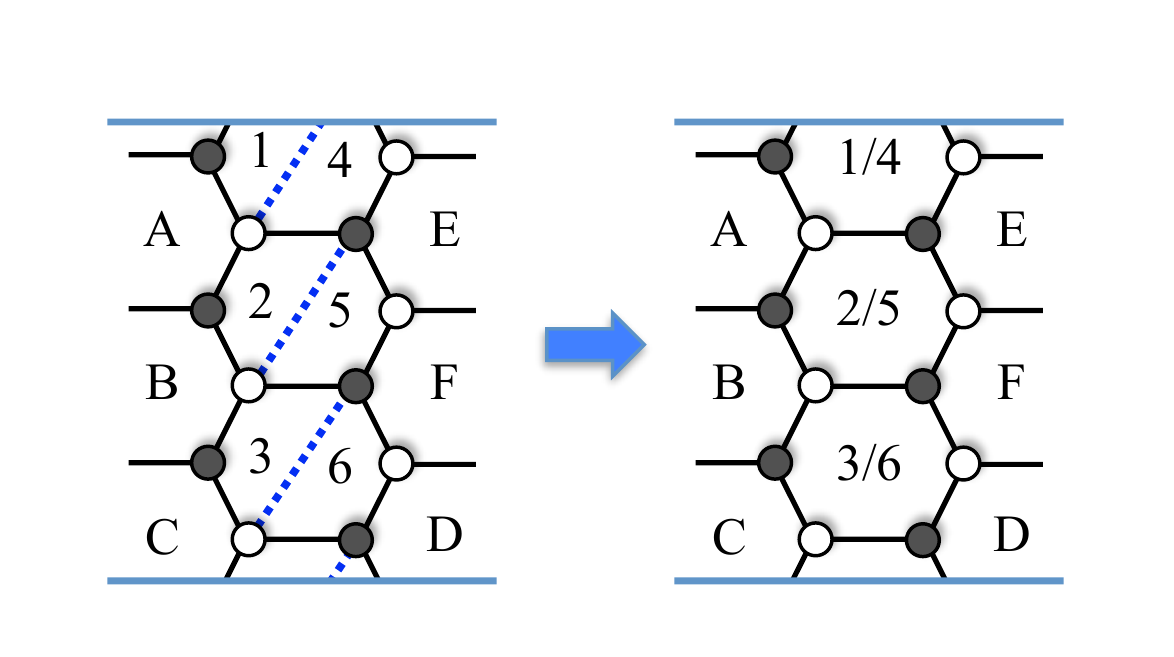}
\vspace{-.3cm}\caption{Higgsing in the dimer corresponding to removal of a puncture separating two sectors with $\NN=1$ gluing.}
\label{sqsq}
\end{center}
\end{figure}

Finally, consider the case of a puncture separating two columns of squares, see Figure \ref{sqsq}. The Higgs mechanism again corresponds to vevs of alternating edges, as shown in the figure. The nodes become cubic, and adjacent squares fuse into hexagons. The process trades two columns of squares by one column of hexagons, hence closing one puncture, and adequately modifying two $\NN=1$ gluings into one $\NN=2$ gluing.

The higgsings we discussed above require introducing vevs in an invariant fashion under the cyclic $\IZ_k$ symmetry of the theory. However, there are simpler choices of higgsing vevs which achieve the closure of a minimal puncture; they break the cyclic symmetry so they correspond to new representatives of ${\cal S}_k$ class theories which are labeled by an additional discrete charge. In fact, given the relation in \sref{sec:global-symm} between minimal punctures and vertical zig-zag paths, it is clear that we can close a minimal puncture by introducing a vev that removes at least one edge in the corresponding path in the dimer. This kind of baryonic higgsing has been considered in \cite{Gaiotto:2015usa}, and leads to a breaking of the $\IZ_k$ orbifold symmetry, namely leads to a whole class of ${\cal S}_k$ theories labeled by some discrete charge. 

\begin{figure}[h]
\begin{center}
\includegraphics[width=14cm]{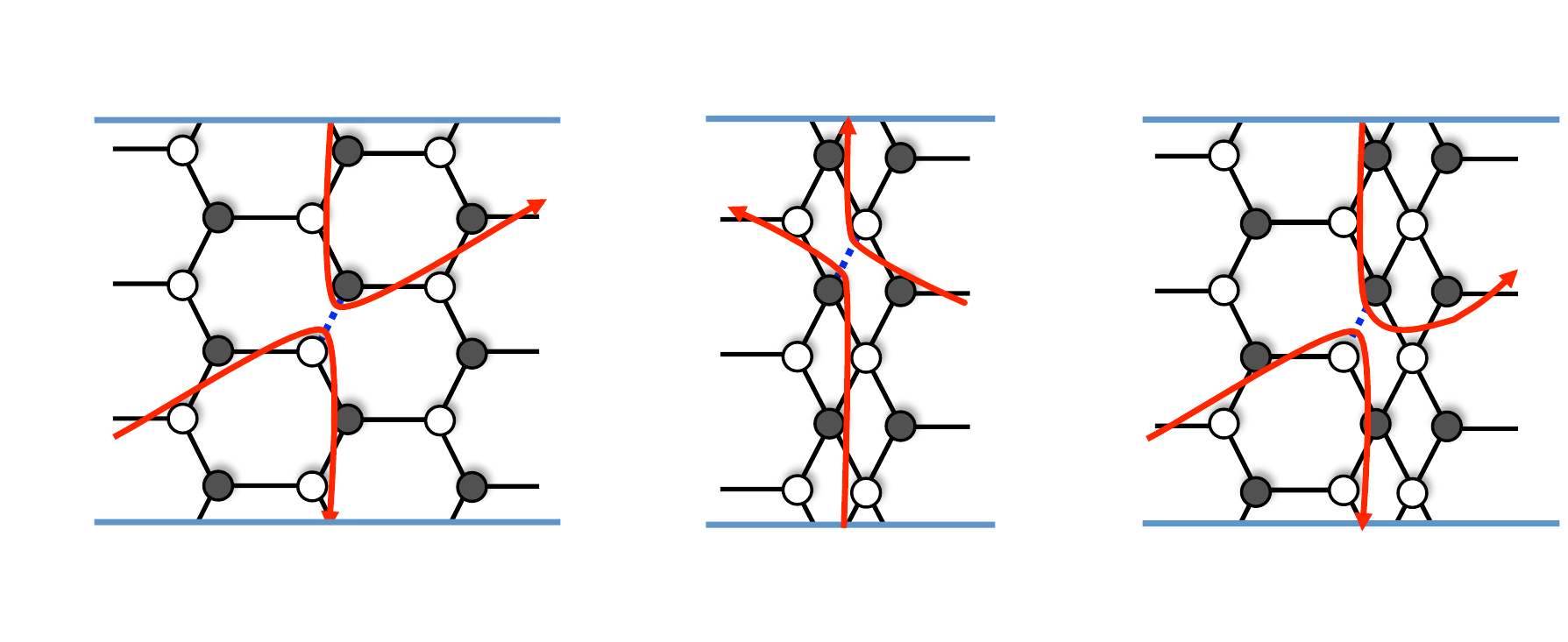}
\vspace{-.7cm}\caption{Three examples of higgsing in the dimer corresponding to closing a minimal puncture.  We show the recombined zig-zag paths that capture the spontaneous symmetry breaking of the global symmetry.}
\label{higgs-discrete}
\end{center}
\end{figure}

This type of higgsing is illustrated in \fref{higgs-discrete} for several examples obtained by $\NN=2$ or  $\NN=1$ gluing. For simplicity we have not integrated out fields entering 2-valent nodes of the superpotential. The baryonic vev removes one edge from the dimer, causing the recombination of the original minimal puncture zig-zag path and one of the intrinsic global symmetry zig-zag paths. This process precisely captures the spontaneous breaking of the two global $U(1)$ symmetries under which the field receiving a vev is charged down to the diagonal combination. In these examples the recombined path is open and is interpreted as a redefined intrinsic global symmetry path. The disappearance of the vertical path corresponds to the closure of the minimal puncture.

The resulting theories are still described by bipartite graphs, illustrating the power of BFTs to capture theories in the general ${\cal S}_k$ class. Whether BFTs suffice to describe all the weak coupling regimes of class ${\cal S}_k$ theories is an interesting question.

\section{The Superconformal Index}
\label{sec:index}

In this section we describe the computation of the superconformal index \cite{Romelsberger:2005eg, Kinney:2005ej} of this class of theories  (see appendix \ref{app:sci} for background). The superconformal index is a powerful tool to understand various properties of superconformal field theories. It has been used for example to check S-dualities among theories of class $\mathcal{S}$ \cite{Gadde:2009kb}, and one can even compute the index of certain class $\mathcal{S}$ theories which do not admit a Lagrangian descirption \cite{Gadde:2010te, Gaiotto:2012xa} (see \cite{Rastelli:2014jja}, for a review). It also gives a nice check for Seiberg duality \cite{Romelsberger:2007ec, Dolan:2008qi}. In \cite{Gaiotto:2015usa} the computation was extended to a large class of theories of class $\mathcal{S}_k$. In this section, we compute the superconformal index in our core ${\cal S}^1_k$ theories, by exploiting systematically the gluing of trinions in the two possible ways introduced in \sref{sec:gluing-maximal}. The existence of two different kinds of minimal punctures, equivalently of two different gluing prescriptions, produces a more intricate duality properties of the superconformal indices, arising from the Seiberg duality in \sref{sec:seiberg}.

\bigskip

\subsection{Gluing Free Trinions}

\begin{figure}[h]
\begin{center}
\includegraphics[width=8cm]{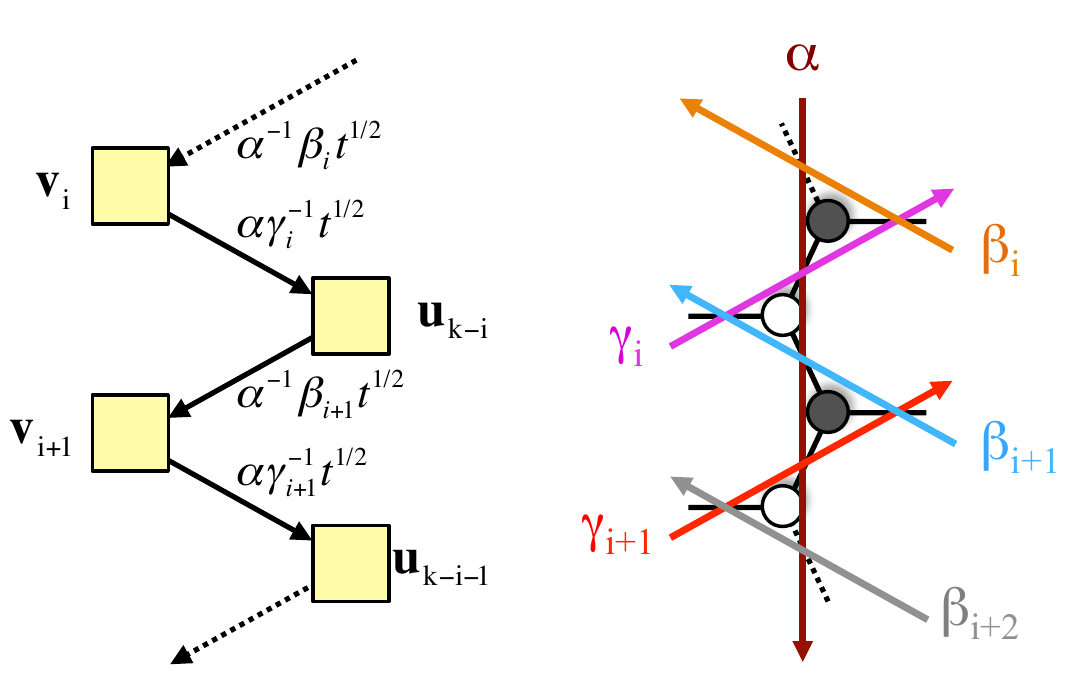}
\end{center}
\vspace{-.35cm}\caption{Left: quiver diagram for a piece of the free trinion. Dashed arrows for non-dynamical fields have been omitted in order to avoid clutter. Right: the corresponding dimer and the zig-zag path associated to the different fugacities.
}
\label{fig:free.trinion}
\end{figure}

To compute the superconformal index of a trinion, we introduce fugacities associated to the global symmetries described in \sref{sec:global-symm}, see \fref{fig:free.trinion}. We denote by $\alpha$ the fugacity for the minimal puncture $U(1)$ (associated to the vertical zig-zag path); for the intrinsic $U(1)^{k-1}\times U(1)^{k-1}\times U(1)$ we use $\beta_i$, $\gamma_i$, and $t$, with $i=1,\ldots, k$ and $\prod_i \beta_i=1$, $\prod_i \gamma_i=1$; and we introduce ${\bf u}_i$, ${\bf v}_i$ (each of which corresponds to an $N$-tuple worth of fugacities) for the global $SU(N)_i$, ${\widetilde {SU(N)}}_i$, $i=1,\ldots, k$ at the maximal punctures. 
%

The basic structure of the index and its behavior under gluing can be illustrated by considering the simple case of $N=2$. For simplicity let us also take $k=2$, the general $k$ case being very similar. 
 The matter content of the free trinion with $N=2, k=2$ is four chiral multiplets in  bifundamental representations of the maximal puncture symmetry factors. Since it is a free theory, the index can be easily computed using \eqref{index.chiral} \cite{Gaiotto:2015usa}
\begin{eqnarray}
\mathcal{I}_{ft}({\bf v}_i, \alpha, {\bf u}_i; \beta, \gamma, t)&:=&\Gamma_e\left((pq)^{\frac{r_1}{2}}t^{\frac{1}{2}}\beta\alpha^{-1}v_1^{\pm 1}u_2^{\pm 1}\right)\Gamma_e\left((pq)^{\frac{\tilde{r}_1}{2}}t^{\frac{1}{2}}\gamma^{-1}\alpha v_1^{\pm 1}u_1^{\pm 1}\right)\nonumber\\
&&\Gamma_e\left((pq)^{\frac{r_2}{2}}t^{\frac{1}{2}}\beta^{-1}\alpha^{-1}v_2^{\pm 1}u_1^{\pm 1}\right)\Gamma_e\left((pq)^{\frac{\tilde{r}_2}{2}}t^{\frac{1}{2}}\gamma\alpha v_2^{\pm 1}u_2^{\pm 1}\right), \nonumber \\
\label{free.trinion}
\end{eqnarray}
where we use the standard notation  $\Gamma_e(z^{\pm 1}):= \Gamma_e(z)\Gamma_e(z^{-1})$, etc. Note that the subindex is always understood as modulo $2$. Also, 
we have allowed for general R-charges $r_1, r_2, \tilde{r}_1, \tilde{r}_2$ for the four bifundamental chiral multiplets,\footnote{In the theory as it stands, or upon gluing it with ingredients preserving the cyclic symmetry around maximal punctures, these R-charges are related. We however prefer to keep them general, as in general there may be gluings to sectors not preserving the cyclic symmetry, e.g. by a higgsing vev as in \sref{section_closing_minimal_punctures}.} since upon gluing one gets interacting theories whose R-charges are fixed by $a$-maximization \cite{Intriligator:2003jj}, and in general they differ from the free field ones. Let us now indeed proceed to the construction of interacting theories by gluing.

\bigskip

\noindent{\bf $\bullet$ {$\mathcal{N}=2$ Gluing}}

\medskip

\begin{figure}[h]
\begin{center}
\includegraphics[width=11cm]{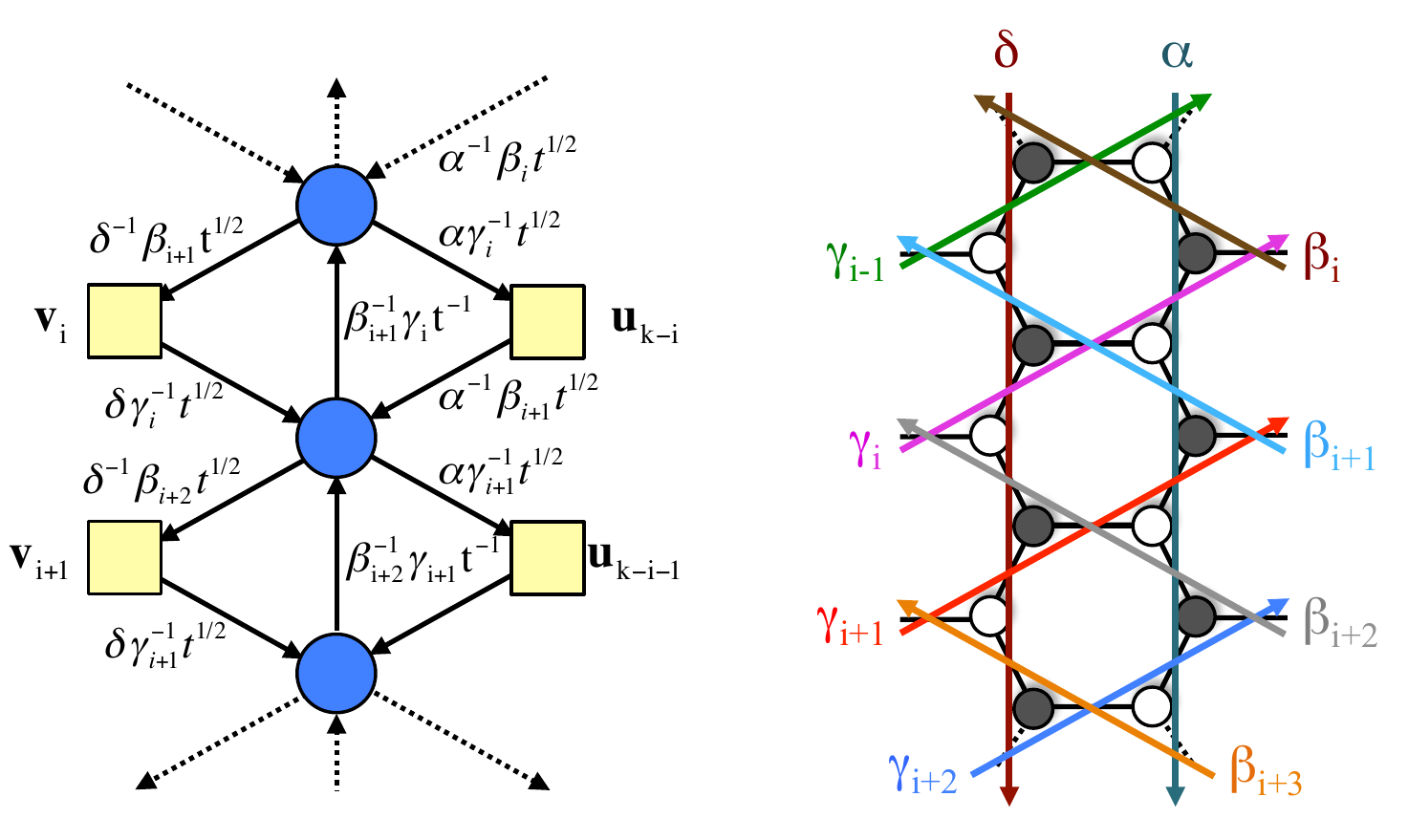}
\end{center}
\vspace{-.35cm}\caption{Left: quiver diagram for a piece of the basic interacting theory corresponding to the four-punctured sphere obtained by the $\mathcal{N}=2$ gluing of two free trinions. Right: the corresponding dimer and the zig-zag paths associated to the different fugacities.}
\label{fig:N2gluing}
\end{figure}

Let us first glue two free trinions by the $\mathcal{N}=2$ gluing, essentially as in \cite{Gaiotto:2015usa}. This is done by gauging the maximal puncture global symmetry, and introducing bifundamental chiral multiplets of the gauged factor, as well as cubic superpotentials. This is described by the quiver/dimer diagrams in \fref{fig:N2gluing}.

The index of the theory is 
\begin{eqnarray}
\mathcal{I}_{{\bf v} \delta\alpha{\bf u}} &=& \frac{[(q;q)(p;p)]^2}{4}\oint\frac{dz_1}{2\pi iz_1}\oint\frac{dz_2}{2\pi i z_2}\frac{1}{\Gamma_e(z_1^{\pm 2})\Gamma_e(z_2^{\pm 2})}\nonumber\\
&&\times\;\Gamma_e\left((pq)^{\frac{r}{2}}t^{-1}\beta\gamma z_1^{\pm 1}z_2^{\pm 1}\right)\Gamma_e\left((pq)^{\frac{\tilde{r}}{2}}t^{-1}(\beta\gamma)^{-1} z_1^{\pm 1}z_2^{\pm 1}\right)\;\nonumber\\
&&\times\;  \mathcal{I}_{ft}({\bf v}_i,\delta, {\bf z}_{1-i}; \beta^{-1}, \gamma, t)\; \mathcal{I}_{ft}( {\bf z}, \alpha, {\bf u}; \beta, \gamma, t),
\label{4punctures1}
\end{eqnarray} 
where the first line represents the two $SU(2)$ gaugings, and the second line represents the contribution of the bifundamental chiral multiplets.  Notice that the R-charges are constrained by  the presence of the superpotential terms described in the quiver/dimer.
For general $k$, $\beta^{-1}$ should be interpreted more formally as $\beta_{i+1}$. Similarly, if the gluing to the trinion in \fref{fig:free.trinion} in done by the left (namely gauging the fugacity ${\bf u}$ in \eqref{free.trinion}), we need to shift $\beta_i$ to $\beta_{i-1}$.

The expression \eqref{4punctures1} is the index for a basic interacting theory or a basic four-punctured sphere \cite{Gaiotto:2015usa} with a general assignment of R-charges.  
Let us introduce the notation
\beqa
\mathcal{I}_{{\bf v}\delta\alpha{\bf u}} \equiv \int [d{\bf z}]_{\NN=2} \,\mathcal{I}_{ft}({\bf v}_i,\delta, {\bf z}_{k-i+1}; \beta_{i+1}, \gamma_i, t)\; \mathcal{I}_{ft}( {\bf z}_i, \alpha, {\bf u}_i; \beta_i, \gamma_i, t),\label{N2glue}
\eeqa
where the diverse ingredients in the $\NN=2$ gluing, in particular the bifundamental chiral multiplets among groups in the column, are implicit in the `measure' $[d{\bf z}]_{\NN=2}$.

\bigskip

\noindent{\bf $\bullet$ {$\mathcal{N}=1$ Gluing}}

\medskip

\begin{figure}[h]
\begin{center}
\includegraphics[width=8.5cm]{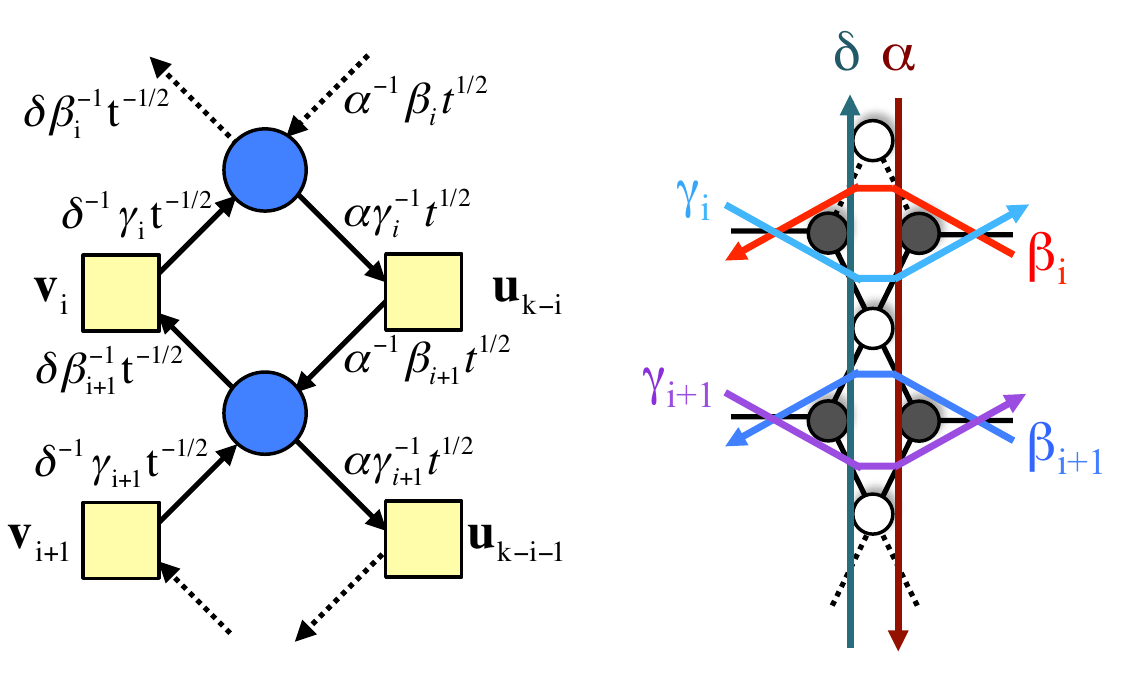}
\end{center}
\vspace{-.35cm}\caption{Left: quiver diagram for a piece of the basic interacting theory corresponding to the four-punctured sphere obtained by the $\mathcal{N}=1$ gluing of two free trinions. Right: the corresponding dimer and the zig-zag paths associated to the different fugacities.}
\label{fig:N1gluing}
\end{figure}

The supersymmetric index after the $\mathcal{N}=1$ gluing is
\begin{eqnarray}
\tilde{\mathcal{I}}_{{\bf v}\delta\alpha{\bf u}} &=& \frac{[(q;q)(p;p)]^2}{4}\oint\frac{dz_1}{2\pi iz_1}\oint\frac{dz_2}{2\pi i z_2}\frac{1}{\Gamma_e(z_1^{\pm 2})\Gamma_e(z_2^{\pm 2})}\nonumber\\
&&\times\; \mathcal{I}_{ft}({\bf z}_i^{-1}, \delta^{-1}, {\bf v}_{i}^{ -1}; \beta^{-1}, \gamma^{-1}, t^{-1})\; \mathcal{I}_{ft}({\bf z},\alpha, {\bf u}; \beta, \gamma, t).
 \label{4punctures2}
\end{eqnarray}
In this case the gluing of fugacities does not require to shift $\gamma_i$ or $\beta_i$, in contrast with the $\mathcal{N}=2$ gluing. A non-trivial operation for the gluing is that we need to perform the charge conjugation for all the fields in one of the free trinions (this is the necessary flipping of the colors of the vertices in the dimer we described in \sref{sec:gluing-maximal}). 
This corresponds to inverting the directions of the zig-zag paths as well as the arrows representing chiral multiplets. 

The expression \eqref{4punctures2} is the index for another basic interacting theory, i.e. for another basic four-punctured sphere. 
Let us introduce the notation
\beqa
\tilde{\mathcal{I}}_{{\bf v}\delta\alpha{\bf u}} \equiv \int [d{\bf z}]_{\NN=1} \,\mathcal{I}_{ft}({\bf v}_i^{-1},\delta, {\bf z}_{k-i+1}^{-1}; \gamma_{i}, \beta_{i+1}, t^{-1})\; \mathcal{I}_{ft}( {\bf z}_i, \alpha, {\bf u}_i; \beta_i, \gamma_i, t). \label{N1glue}
\eeqa
Note that in this case there are no bifundamental chiral multiplets among gauge groups in the column. The equivalence between \eqref{N1glue} with $k=2$ and \eqref{4punctures2} can be understood by using the identity of the free trinion
\begin{equation}
\mathcal{I}_{ft}({\bf v}_i^{-1},\delta, {\bf z}_{k-i+1}^{-1}; \gamma_{i}, \beta_{i+1}, t^{-1}) = \mathcal{I}_{ft}({\bf z}_i^{-1},\delta^{-1}, {\bf v}_{k-i}^{-1}; \beta_{i}^{-1}, \gamma_{i}^{-1}, t^{-1}).
\end{equation}

It is straightforward to extend the procedure to glue more trinions and construct the arbitrary core theories of \sref{sec:general-class}.

\bigskip

\subsection{Dualities}
\label{sec:index-dual}

One advantage of the description of 4d gauge theories in terms of Riemann surfaces (on which suitable 6d theories are compactified) is that S-dualities become geometrized in terms of exchange of punctures. This leads to interesting invariance properties of the index under such operations.
For theories in the $\mathcal{N}=2$ class $\mathcal{S}$ or a family of $\mathcal{N}=1$ class $\mathcal{S}_k$ theories, the exchange between punctures of the same type has been studied in \cite{Gaiotto:2009we,Gaiotto:2015usa}; for the $\NN=1$ class $\mathcal{S}$, its maximal punctures may or may not carry mesons, and their exchange (or equivalently the exchange of two minimal punctures as well as of the attached two-spheres) is related to Seiberg duality. This is in complete analogy with the realization of Seiberg duality by crossing of NS- and NS'-branes in \cite{Elitzur:1997fh}. For our class of ${\cal S}^1_k$ theories the analogous phenomenon has been discussed in \sref{sec:seiberg}. In this section we revisit the result from the perspective of the index, and describe its transformation properties under exchange of punctures.

Clearly, exchange of identical punctures leads to same results as in \cite{Gaiotto:2015usa}. Namely, the index \eqref{4punctures1} of the gauge theory obtained from two free trinions with $\NN=2$ gluing has the invariances
\begin{equation}
\mathcal{I}_{{\bf v}\delta\alpha{\bf u}} = \mathcal{I}_{{\bf v}\alpha \delta {\bf u}} = \mathcal{I}_{{\bf u}\delta\alpha{\bf v}}.
\end{equation}
Namely, the index is invariant under the exchange of punctures of the same type. The same result holds if this theory is glued to additional sectors, to its left and its right, with 
$\NN=2$ gluing.

\begin{figure}[h]
\begin{center}
\includegraphics[width=12cm]{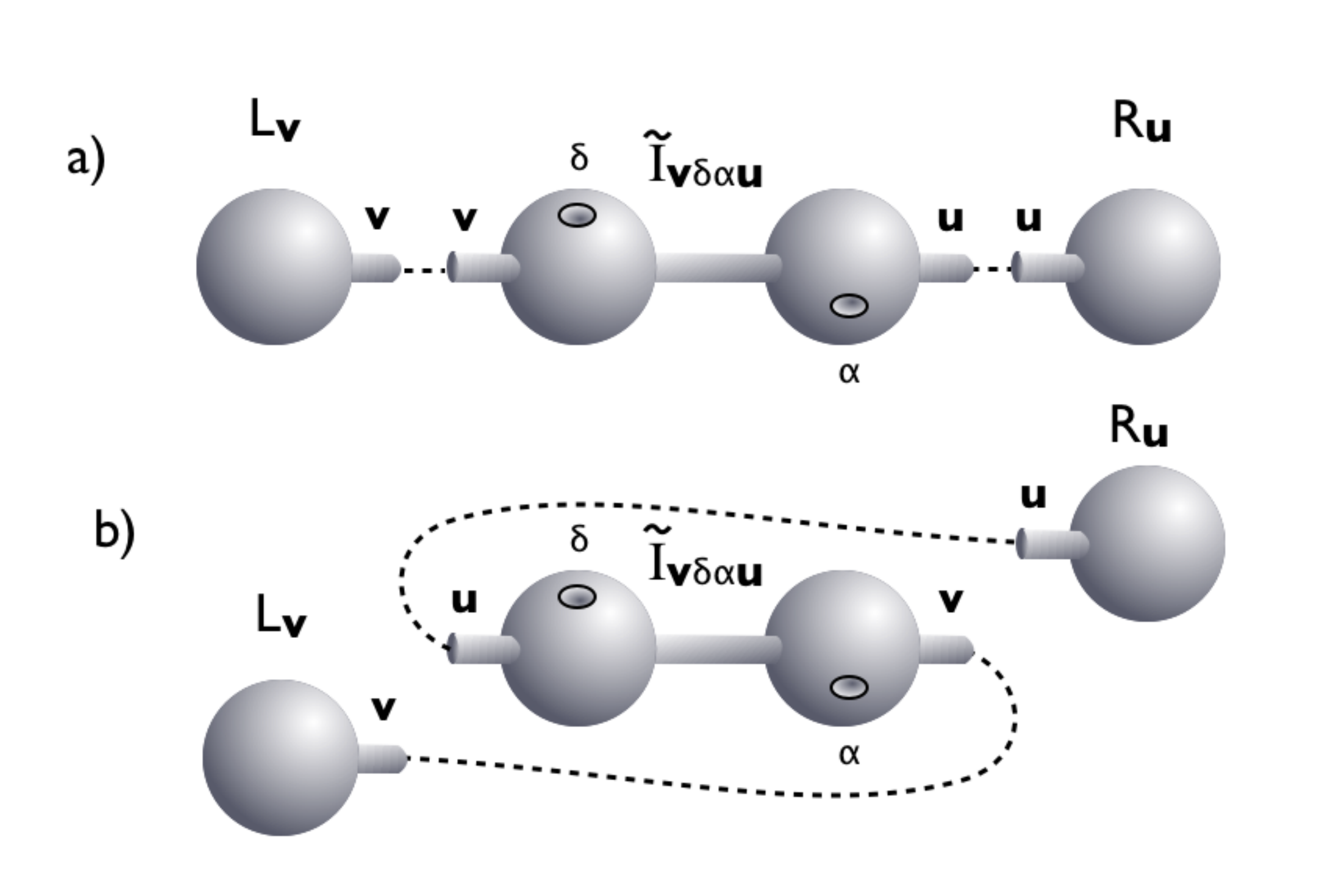}
\caption{a) General theory. b) Seiberg duality from exchange of maximal punctures.}
\label{seiberg-riemann}
\end{center}
\end{figure}

Let us turn to the more interesting case involving $\mathcal{N}=1$ gluing, which may yield different kind of punctures. 
The simplest case corresponds to the theory obtained from two free trinions with $\mathcal{N}=1$ gluing. Its index is given by \eqref{4punctures2}, corresponding to the gauge factors and bifundamental chiral multiplets, in a column of squares in the dimer, recall \fref{orbifold-nodes}. In general we will be interested in studing the properties of this sector in situations in which it is glued to arbitrary Riemann surfaces both on the left and right, with indices $L_{\bf v}$, $R_{\bf u}$, see \fref{seiberg-riemann}.a. This gluing can be taken to be of $\NN=2$ or $\NN=1$ kind on either side. To illustrate the main properties, it suffices to focus on one example with one gluing of each kind, say
\beqa
\int [d{\bf v}]_{\NN=2}\, [d{\bf u}]_{\NN=1}\, L_{\bf v} {\tilde I}_{{\bf v}\delta\alpha{\bf u}}\, R_{\bf u},
\label{complete-th}
\eeqa
where any shift of subindex or inversion of fugacities associated to the $\mathcal{N}=2, 1$ gluings are implicit for notational simplicity. In practice, one can use the explicit formulae of \eqref{N2glue} and \eqref{N1glue}. We are interested in the behavior of this quantity upon the application of Seiberg duality in the middle block. Naively, this would seem to correspond to the exchange of the minimal punctures, but that does not actually correspond the exchange between an NS-brane and an NS'-brane in the Type IIA setup. This transformation between the minimal punctures will be considered towards the end of this section.

Actually, as it is also the case for the $\NN=1$ class ${\cal S}$ theories, Seiberg duality corresponds to the exchange of maximal punctures, which effectively implements the exchange of the NS- and NS'-branes (including the information on the kind of minimal puncture), see Figure \ref{seiberg-riemann}.b. By applying \eqref{rains} to \eqref{4punctures2}, we obtain\footnote{In \eqref{sym2}, ${\bf u}^{-1}_i$ and ${\bf v}^{-1}_i$ after the duality are more precisely ${\bf u}^{-1}_{k-i}$  and ${\bf v}^{-1}_{k-i}$ for general $k$. but the difference is irrelevant for $k=2$.} 
\begin{eqnarray}
\tilde{\mathcal{I}}_{{\bf v}\delta\alpha{\bf u}}(\beta, \gamma, t) &=& \tilde{\mathcal{I}}_{{\bf u}^{-1} \delta^{-1}\alpha^{-1}{\bf v}^{-1}}(\gamma^{-1}, \beta^{-1}, t) \nonumber\\
&&\Gamma_e \Big( (pq)^{\frac{r'_1+\tilde{r}'_1}{2}}t^{-1}\beta^{-1}\gamma v_1^{\pm 1}v_2^{\pm 1}\Big)\; \Gamma_e\Big((pq)^{\frac{r'_2+\tilde{r}'_2}{2}}t^{-1}\beta\gamma^{-1} v_1^{\pm 1}v_2^{\pm 1}\Big)\nonumber\\
&&\Gamma_e\left((pq)^{\frac{r_1+\tilde{r}_1}{2}}t\beta\gamma^{-1}u_1^{\pm 1}u_2^{\pm 1}\right)\; \Gamma_e\left((pq)^{\frac{r_2+\tilde{r}_2}{2}}t\beta^{-1}\gamma u_1^{\pm 1}u_2^{\pm 1}\right), \label{sym2}
\end{eqnarray}
The R-charges of the index before and after the duality are related by $r'_i \leftrightarrow \tilde{r}'_i$ and $r_i\leftrightarrow \tilde{r}_i$ for $i=1, 2$. The exchange of the punctures produces the appearance of extra degrees of freedom, corresponding to the mesons of the dualized gauge factors, transforming as bifundamentals of the flavour groups at the maximal punctures $SU(2)_1 \times SU(2)_2$ and $\widetilde{SU(2)}_1 \times \widetilde{SU(2)}_2$. In complete theories like (\ref{complete-th}), any $\NN=2$ gluing implicitly contains bifundamentals in the measure $[d\ldots]_{\NN=2}$, whose contribution cancels precisely against these mesons. The mesons disappear and simultaneously the gluing is turned into $\NN=1$; this nicely dovetails the field theory and dimer analysis in \sref{sec:seiberg}. Similarly, for any $\NN=1$ gluing the measure $[d\ldots]_{\NN=1}$ combines with the mesons to produce an $\NN=2$ gluing measure $[d\ldots]_{\NN=2}$, again in complete agreement with the field theory/dimer description. For \eqref{complete-th}, we schematically have
\beqa
\int [d{\bf v}]_{\NN=2}\, [d{\bf u}]_{\NN=1}\, L_{\bf v} {\tilde I}_{{\bf v}\delta\alpha{\bf u}}\, R_{\bf u} =\int [d{\bf v}]_{\NN=1}\, [d{\bf u}]_{\NN=2}\, L_{\bf v}{\tilde I}_{{\bf u}^{-1} \delta^{-1}\alpha^{-1}{\bf v}^{-1}}\, R_{\bf u} .\label{index.seiberg}
\eeqa
This expresses the invariance of the index under Seiberg duality.

In writing \eqref{index.seiberg}. we have only written explicitly the fugacities associated to the punctures. In fact, the relation \eqref{sym2} also ensures the consistency of the fugacities associated to the intrinsic flavor symmetries. Note that the fugacities associated to $U(1)_{\beta} \times U(1)_{\gamma}$ are flipped in \eqref{sym2}. This nicely fits the configuration of zig-zag paths after the duality. Namely, the zig-zag paths for $\beta$ and $\gamma$ of $L_{\bf v}$ and $R_{\bf u}$ automatically connect to the corresponding zig-zag paths for $\beta, \gamma$ of ${\tilde I}_{{\bf u}^{-1}\delta^{-1}\alpha^{-1}{\bf v}^{-1}}$ according to the $\mathcal{N}=1, 2$ gluings, respectively.  Furthermore, the conditions that R-charges have to satisfy after the duality are also automatically guaranteed, given that the R-charges before the duality are assigned consistently.

For completeness, we conclude by discussing the transformation properties of \eqref{4punctures1} under exchange of minimal punctures. By applying the symmetry \eqref{E7} to \eqref{4punctures2}, 
it is possible to show that 
\begin{eqnarray}
&&\tilde{\mathcal{I}}_{{\bf v}\delta\alpha{\bf u}} = \tilde{\mathcal{I}}_{{\bf v}\alpha\delta{\bf u}}\;
\Gamma_e\left((pq)^{r_1}\beta^{2}\alpha^{-2}t \right)\Gamma_e\left((pq)^{\tilde{r}_1}\gamma^{-2}\alpha^{2}t\right)
\Gamma_e\left((pq)^{r_2}\beta^{-2}\alpha^{-2}t\right)\Gamma_e\left((pq)^{\tilde{r}_2}\gamma^{2}\alpha^{2}t\right)
\nonumber \\
&&\quad \Gamma_e\left((pq)^{r_1'}\beta^{-2}\delta^{2}t^{-1}\right)\Gamma_e\left((pq)^{\tilde{r}_1'}\gamma^2\delta^{-2}t^{-1}\right)
\Gamma_e\left((pq)^{r'_2}\beta^{2}\delta^{2}t^{-1}\right)\Gamma_e\left((pq)^{\tilde{r}'_2}\gamma^{-2}\delta^{-2}t^{-1}\right)
\nonumber
,
\end{eqnarray}
for general assignments of R-charges (consistent with the superpotential couplings). The R-charges of the index before and after the duality are related by $r'_i \leftrightarrow 1 - r_i$ and $\tilde{r}'_i \leftrightarrow 1 - \tilde{r}_i$ for $i=1, 2$. Hence, the index \eqref{4punctures2} is invariant under the exchange of the two minimal punctures up to contributions from `baryons' (gauge singlets charged under the minimal puncture $U(1)$'s), which are necessary for `t Hooft anomaly matching. This can be regarded as a new kind of duality property of these theories. It would be interesting to gain further physical intuition about it, beyond its realization as a transformation of the index.

\bigskip

\section{Conclusions}

\label{section_conclusions}

We undertook a thorough investigation of new features of the class ${\cal S}_k$ of 4d $\NN=1$ SCFTs, which are constructed as compactifications of the 6d $(1,0)_k$ SCFTs on punctured Riemann surfaces. To do so, we introduced and studied a large family of quiver theories in class ${\cal S}_k$, which we dubbed class ${\cal S}^1_k$, largely extending ideas in \cite{Gaiotto:2015usa}. These theories have a Type IIA realization as $\IZ_k$ orbifolds of configurations of D4-branes stretched among relatively rotated NS- and NS'-branes. Interestingly, all examples currently known of quiver theories in class ${\cal S}_k$ (including ours and those in \cite{Gaiotto:2015usa}) are examples of Bipartite Field Theories.

Taking into account the non-trivial anomalous dimensions that these theories generically possess, we established the correspondence between the complex structure parameters of the underlying Riemann surfaces, more precisely the positions of minimal punctures (minus one), and the marginal couplings of the theories.

The full ${\cal S}^1_k$ class can be constructed by gluing basic building blocks, given by the free trinions introduced in \cite{Gaiotto:2015usa}. There are two different kinds of gluings, which we denoted $\NN=2$ and $\NN=1$, according to their supersymmetry structure in the parent theory before orbifolding. 

We described the dualities in this class of theories as exchanges of punctures, and studied the transformations that correspond to Seiberg dualities. We also described the computation of the superconformal index, and used it to check the duality invariance.

A natural question is how to go beyond the core theories investigated in this article and \cite{Gaiotto:2015usa}, which correspond to a sphere with two maximal punctures and a number of minimal punctures. In particular, it would be desirable to understand whether it is possible to increase the number of maximal punctures or the genus of the compactification Riemann surface. To achieve this goal, a crucial step is the construction of the theory corresponding to a sphere with three maximal punctures. This theory is presumably severely constrained by its global symmetries and consistency with the theories with two maximal punctures upon closure of some of the maximal punctures, but the possible absence of a weakly coupled description makes its formulation challenging. Preliminary progress in this direction has been achieved in \cite{Gaiotto:2015usa}. We leave this interesting question for future work.

A more modest, yet interesting, problem is to carry out a systematic study of the construction of theories on a 2-torus by gluing the two maximal punctures in core ${\cal S}^1_k$ theories. This results in the class of SCFTs associated to D3-branes probing toric Calabi-Yau 3-folds. Indeed, considering an appropriate initial ${\cal S}^1_k$ theory and general higgsings, it is possible to reach the SCFT for any toric Calabi-Yau 3-fold. This implies that there is an interesting web of relations with the SCFTs of D3-branes at singularities, and presumably to their gravity realizations in the context of the AdS/CFT correspondence. We expect the theories defined on a torus present several novelties in their structure of minimal punctures and counting of marginal couplings, which certainly deserve further study. In addition, we generally expect different possible 6d realizations of the same torus theories. It would thus be interesting to understand in detail what the connection between such theories is. 

The connection with D3-branes at orbifolds also raises prospect of finding novel realizations of the deconstruction proposal in \cite{ArkaniHamed:2001ie}. This would connect with the viewpoint advocated in \cite{Franco:2013ana}, which conjectures that certain limits of BFTs (in suitable large numbers of gauge factors) deconstruct higher-dimensional field theories.

We hope return to these and other open questions in future work.

\bigskip

\section*{Acknowledgments}

We would like to thank S. Razamat for very useful discussions. The work of H. H. and A. U.  is supported by the Spanish Ministry of Economy and Competitiveness under grants FPA2012-32828, Consolider-CPAN (CSD2007-00042),  the grants  SEV-2012-0249 of the Centro de Excelencia Severo Ochoa Programme, SPLE Advanced Grant under contract ERC-2012-ADG$\_$20120216-320421 and the contract ``UNILHC" PITN-GA-2009-237920 of the European Commission. The work of H. H. is supported also by the REA grant agreement PCIG10-GA-2011-304023 from the People Programme of FP7 (Marie Curie Action).


\newpage

\appendix

\section{Superconformal Index Basics}
\label{app:sci}

\subsection{Preliminaries}

The $\mathcal{N}=1$ supersymmetric index \cite{Romelsberger:2005eg, Kinney:2005ej}  of four-dimensional supersymmetric theories can be defined as the Witten index of the theory compactified on $\IS^3$ twisted by fugacities $p, q$ associated to the $SU(2)^2$ isometry group, and $\{u_a\}$ associated to global symmetry group $\mathfrak{F}$. Equivalently,  it is defined as the corresponding partition function on $\IS^1 \times \IS^3$. Its structure reads
\begin{equation}
\mathcal{I}(p, q; {\bf u}) = \text{Tr}_{\mathcal{H}_{S^3}}(-1)^Fe^{-\beta\{Q, Q^{\dagger}\}}p^{j_1+j_2-\frac{1}{2}R}q^{j_1-j_2-\frac{1}{2}R}\prod_{a \in \mathfrak{F}}u_a^{q_a}. \label{index}
\end{equation}
Here the trace is taken over the Hilbert space on $\IS^3$, and $F$ is the fermion number. The Hamiltonian is realized in terms of supercharges $Q, Q^{\dagger}$,  which commute with the global symmetries. The quantum number under the (Cartan generators of the) latter are denoted by $j_1, j_2$ (for the  $SU(2)^2$ isometry), by $R$ for the $U(1)$ R-symmetry, and by $q_a$ for the  flavor group $\mathfrak{F}$.

By the standard argument, only states with $\{Q, Q^{\dagger}\} = 0$ contribute to the index. The index is robust, independent of $\beta$ or other continuous parameters, and remains invariant under the RG flow, so it can be computed from a UV description \cite{Romelsberger:2005eg, Romelsberger:2007ec, Festuccia:2011ws} (assuming that there are no emergent $U(1)$ symmetries in the IR). 

The UV descriptions are in terms of weakly coupled chiral and vector multiplets. The single particle index for a chiral multiplet, and a vector multiplet of a gauge group $G$, are respectively
\beqa
&& f_{\chi}(p, q; u_a) = \frac{(pq)^{\frac{R}{2}}u_a^{q_a} - (pq)^{-\left(\frac{R}{2}-1\right)}u_a^{-q_a}}{(1-p)(1-q)},
\nonumber \\
&& f_{v}(p, q; z_\alpha) = \frac{2pq - (p+q)}{(1-p)(1-q)}\chi_{\text{adj(G)}}({\bf z}), 
\label{single-index}
\eeqa
where $\chi_{\text{adj(G)}}({\bf z})$ represents the character of the adjoint representation of $G$. Here the $z_\alpha$ are fugacities in the Cartan subalgebra of $G$, which is regarded as a global symmetry at this stage.

Their multi-particle indices are simply obtained by applying to \eqref{single-index} the plethystic exponential   
\begin{equation}
\mathcal{I}_f = \exp\left(\sum_{k=1}^{\infty} \frac{f(p^k, q^k;u^k)}{k}\right).
\end{equation}
Hence, the supersymmetic index of the chiral multiplet is
\begin{equation}
\mathcal{I}_{\chi}(p, q; u_a) = \prod_{i,j=0}^{\infty}\frac{1-p^{1-\frac{R}{2}+i}q^{1-\frac{R}{2}+j}u_a^{-q_a}}{1-p^{\frac{R}{2}+i}q^{\frac{R}{2}+j}u_a^{q_a}} =: \Gamma_e\left((pq)^{\frac{R}{2}}u_a^{q_a}\right), \label{index.chiral}
\end{equation}
where we used the standard notation of the elliptic Gamma function $\Gamma_e(z)$. 

The multi-particle index of e.g. an $SU(N)$ vector multiplet (easily generalized to other groups) is 
\begin{equation}
\mathcal{I}_{SU(N)}(p, q; {\bf z})  = \left\{(q;q)(p;p)\right\}^{N-1}\prod_{\alpha \neq \beta}^N\frac{1}{\left(1 - \frac{z_\alpha}{z_\beta}\right)\Gamma_e\left(\frac{z_\alpha}{z_\beta}\right)}, \label{index.SUN}
\end{equation}
where we defined the index for an abelian vector multiplet
\begin{equation}
 (q;q)(p;p) := \prod_{\alpha=0}^{\infty}\left(1 - q^{\alpha+1}\right) \prod_{\beta=0}^{\infty}\left(1 - p^{\beta+1}\right) .
\end{equation}
 Note that for $SU(N)$ we have $\prod_{\alpha=1}^Nz_\alpha = 1$. 

When a theory $\mathcal{T}$ has a flavor symmetry $G$,  we can gauge it to obtain a new theory $\mathcal{T}_G$. The index of the latter is simply obtained by multiplying the original index $\mathcal{I}_{\mathcal{T}}(z)$ times the vector multiplet index $\mathcal{I}_v(z)$ and integrating over the fugacities $z$ of $G$, with the Haar measure $[dz]_G$
\begin{equation}
\mathcal{I}_{\mathcal{T}_G} = \oint [dz]_G \; \mathcal{I}_{v}(z) \; \mathcal{I}_{\mathcal{T}}(z). \label{gauging}
\end{equation}
For an $SU(N)$ gauging, the Haar measure is 
\begin{equation}
[dz]_{SU(N)} = \frac{1}{N!}\left[\prod_{\alpha=1}^{N-1} \frac{dz_\alpha}{2\pi i z_\alpha}\right]\left[\prod_{\alpha \neq \beta}^N\left(1 - \frac{z_\alpha}{z_\beta}\right) \right],
\end{equation}
with $\prod_{\alpha=1}^N z_{\alpha}=1$. So, using \eqref{index.SUN}, we obtain the index
\begin{equation}
\mathcal{I}_{\mathcal{T}_{SU(N)}} = \frac{\left\{(q;q)(p;p)\right\}^{N-1}}{N!}\oint_{\mathbb{T}^{N-1}}\left[\prod_{\alpha=1}^{N-1} \frac{dz_\alpha}{2\pi i z_\alpha}\right]\left[\prod_{i \neq j}^N\Gamma_e\left(\frac{z_\alpha}{z_\beta}\right)\right]^{-1}\mathcal{I}_T({\bf z}), \label{gauging.SUN}
\end{equation}
where $\mathbb{T}$ is the unit circle.

\bigskip

\subsection{Some Useful Formulae}

Here we collect some mathematical results useful to derive the duality formulae in \sref{sec:index-dual}.

Theorem 4.1 in \cite{MR2630038} states that
\begin{equation}
I_{A_n}^{(m)}({\bf t};{\bf u};p,q)=\prod_{0\leq r,s < m+n+2}\Gamma_e(t_ru_s)\;I_{A_m}^{(n)}\left(T^{\frac{1}{m+1}}{\bf t}^{-1} ; U^{\frac{1}{m+1}}{\bf u}^{-1};p,q\right) 
\label{rains}
\end{equation}
when $\prod_{0 \leq r < m+n+2}t_ru_r=(pq)^{m+1}$. Here $T:=\prod_{0\leq r < n+m+2}t_r$, $U:=\prod_{0 \leq r <n+m+2}u_r$, and ${\bf t}^{-1}=(t_0^{-1},\ldots, t_{m+n+1}^{-1})$, and we also defined
\begin{eqnarray}
&&I_{A_n}^{(m)}(t_0, \cdots, t_{m+n+1};u_0,\cdots, u_{n+m+1};p,q)\nonumber\\
&&:=\frac{(p;p)^n(q;q)^n}{(n+1)!}\oint\prod_{\alpha=1}^n\frac{dz_\alpha}{2\pi iz_\alpha}\frac{\prod_{0\leq \alpha\leq n}\prod_{0\leq r<m+n+2}\Gamma_e(t_rz_\alpha)\Gamma_e(u_r/z_\alpha)}{\prod_{0\leq \alpha < \beta \leq n}\Gamma_e(z_\alpha/z_\beta)\Gamma_e(z_\beta/z_\alpha)}, \nonumber \\
\end{eqnarray}
where $\prod_{\alpha=0}^{n}z_{\alpha} = 1$. 
Another useful formula comes from the elliptic beta integral
\begin{equation}
E^m({\bf t}):=\left(\prod_{0\leq r<s \leq 2m+5}(t_rt_s;p,q)\right)\frac{(p;p)(q;q)}{2}\oint\frac{dz}{2\pi iz}\frac{\prod_{r=0}^{2m+5}\Gamma_e(t_rz^{\pm 1})}{\Gamma_e(z^{\pm 2})}, 
\label{elliptic.beta}
\end{equation}
with $\prod_{r=0}^{2m+5}t_r=(pq)^{m+1}$. We defined
\begin{equation}
(x;p,q) = \prod_{r,s=0}^{\infty}(1 - xp^rq^s).
\end{equation}
When $m=1$, the elliptic beta integral \eqref{elliptic.beta} is invariant under the Weyl group of $E_7$, in particular under the permutation of $t_r$, and also \cite{MR2044635}
\begin{equation}
E^1(t) = E^1(t_0v, t_1v, t_2v, t_3v, t_4/v, t_5/v, t_6/v, t_7/v), 
\label{E7}
\end{equation}
where $v^2 = \frac{pq}{t_0t_1t_2t_3}$. 

\bigskip




\end{document}